\documentclass[aps,pra,twocolumn,amsmath,amssymb,superscriptaddress,showpacs]{revtex4-1}
\usepackage{graphics}
\usepackage{graphicx}
\usepackage{amsfonts}
\usepackage{amssymb}
\usepackage{epsfig}
\usepackage{color}
\usepackage{CJK}
\usepackage{ulem}
\usepackage{multirow}

\begin{document}

\title{Criticality and factorization in the Heisenberg chain \\
       with Dzyaloshinskii-Moriya interaction}

\author {Tian-Cheng Yi}
\affiliation{\mbox{College of Science, Nanjing University of Aeronautics and Astronautics,
Nanjing, 211106, China}}
\affiliation{\mbox{School of Physical Science and Technology, Soochow University, Suzhou,
Jiangsu 215006, China}}

\author {Wen-Long You}
\email{youwenlong@gmail.com}
\affiliation{\mbox{College of Science, Nanjing University of Aeronautics and Astronautics,
Nanjing, 211106, China}}
\affiliation{\mbox{School of Physical Science and Technology, Soochow University, Suzhou,
Jiangsu 215006, China}}

\author {Ning Wu}
\affiliation{\mbox{Center for Quantum Technology Research, School of Physics,
Beijing Institute of Technology, Beijing 100081, China}}

\author {Andrzej M. Ole\'s }
\affiliation{Max Planck Institute for Solid State Research,
             Heisenbergstrasse 1, D-70569 Stuttgart, Germany }
\affiliation{\mbox{Marian Smoluchowski Institute of Physics, Jagiellonian 
         University, Prof. S. \L{}ojasiewicza 11, PL-30348 Krak\'ow, Poland}}

\begin{abstract}
In this work, we address the ground state properties of the anisotropic
spin-1/2 Heisenberg XYZ chain under the interplay of magnetic fields
and the Dzyaloshinskii-Moriya (DM) interaction which we interpret as
an electric field. The identification of the regions of enhanced
sensitivity determines criticality in this model.
We calculate the Wigner-Yanase skew information (WYSI) as a coherence
witness of an arbitrary two-qubit state under specific measurement
bases. The WYSI is demonstrated to be a good indicator for detecting
the quantum phase transitions. The finite-size scaling of coherence
susceptibility is investigated. We find that the factorization line in
the antiferromagnetic phase becomes the factorization volume in the
gapless chiral phase induced by DM interactions, implied by the
vanishing concurrence for a wide range of field. We also present the 
phase diagram of the model with three phases: antiferromagnetic, 
paramagnetic, and chiral, and point out a few common mistakes in 
deriving the correlation functions for the systems with broken 
reflection symmetry.
\end{abstract}

\date{31 May 2019}



\maketitle

\section{Introduction}

Quantum phase transitions (QPTs) that deal with dramatic changes of the
ground state and low-excitation properties induced by small variations
of driving parameters, are one of very active fields of research in
several contexts of modern statistical mechanics, quantum information,
and condensed matter physics. QPTs are believed to take place
exclusively in many-body systems~\cite{Sachdev1999}, while it has been
recently realized that a few-body system may also develop a QPT
\cite{hwang2015prl,hwang2016prl,Liu17,Lar17,wang2018njp}.
Quantum fluctuation accumulated by the non-commutativity between the
driving term and the rest are responsible for the sudden change of the
correlations among the system's constituents.

In view of the central role played in interdisciplinary fields, it is
of crucial importance to devise suitable tools for a proper
characterization of the changes of a quantum system at a QPT. To this
purpose, different quantum-information-based concepts have been put
forward over recent years, in order to identify ground-state variations
across QPTs. Quantum coherence and quantum entanglement are two
characteristic properties of a quantum system. Both of them are
considered to be valuable resources in most quantum information
processing tasks \cite{Aberg2014,Narasimhachar2015,Cwi15,Lo15N,Lo15X}.
Quantum entanglement indicates that a quantum state is nonseparable and
was first pointed out by Schr\"{o}dinger in 1935 for constituent
subsystems \cite{Schrodinger}. Furthermore, entanglement spectrum is
very useful in recognizing certain QPTs in spin systems \cite{You15}.
In contrast, quantum coherence concerns the set of states and is 
usually defined in a given basis by measuring the distance between the 
quantum state $\rho$ and its closest incoherent state for the system 
as a whole~\cite{Baumgratzq14}. Although these two
quantum mechanical properties have a completely different origin,
indications exist that they are equivalent by computing~\cite{Str15}.

It is known that the correlation length tends to be infinite in the
critical regime although the interactions are short-ranged, which can
lead to diverging susceptibilities signalling a QPT. The sensitivity is
greatly enhanced especially for the system at the quantum criticality
comparing with that away from the critical region. In this respect,
quantum-enhanced measurements open the path to many new forms of
enhanced sensitivity across the quantum criticality~\cite{Braun18}.
Taking external fields as probes, the sensitivity given by the
entanglement susceptibility, coherence susceptibility
\cite{Chen16,You17}, and fidelity susceptibility~\cite{Zanardi07,You07}
infers the signatures of quantum critical points and scaling behaviors.
Such strategy is very useful in Hamiltonian engineering, when one
can evaluate the effect of added terms in the Hamiltonian.

To investigate the quantum criticality, we consider a specific
quasi-classical ground state that can be modulated in the anisotropic
quantum antiferromagnetic (AFM) chain under external fields, as pointed
by the pioneering work of Kurmann, Thomas, and M\"{u}ller~\cite{Kur82}.
The separable states are namely the free states in the resource theory
of entanglement~\cite{Horodecki09}. The possible engineering of a
completely separable state is nontrivial and particularly significant
in the presence of strong interactions for information processes
\cite{Vandersypen05} and quantum simulation~\cite{Georgescu14}.
Over recent years, several physical quantities and experimental methods
have been developed for identification and exploration of QPTs.
In the last decade, quantum-information measures, such as in the form
of entanglement and coherence, were found to be an effective tool for
characterizing QPTs and ground-state factorization. Exploring both
criticality and factorization using the tools of quantum information
has proven fruitful in a number of contexts, e.g., low-dimensional spin
models, fermionic systems, cold atom system, and open quantum systems.

In this work, we focus on the one-dimensional (1D) Heisenberg model
with nearest neighbor exchange coupling, which has long served as an
archetype for the study of quantum magnetism in low dimensions.
Strong fluctuations of interacting spins are of particular importance
at low dimension, where the Mermin-Wagner theorem states that thermal
fluctuations prevent long-range order at any finite temperature when
the Hamiltonian obeys a continuous rotational symmetry in spin space.
Even at absolute zero temperature, the zero-point fluctuations may also
prevent long-range order by incorporating additional interactions.
The effects induced by external electric and magnetic
fields have been of particular interest since the magnetic state can
be qualitatively different depending on the magnitude and direction of
the external fields. This has led to interest in the study of QPTs at
finite fields \cite{Brockmann2013,You14,Thakur18}.

Recently, Radhakrishnan, Ermakov, and Byrnes \cite{Rad17} studied 
the quantum coherence in the XY chain with Dzyaloshinsky-Moriya (DM)
interaction and indicated that quantum Jensen-Shannon divergence can
efficiently probe the second-order QPT. Moreover, the local and
intrinsic ingredient among the total quantum coherence can be
discriminated to characterize the first-order QPTs in spin-1/2 XXZ
chain~\cite{Rad16} and the topological QPTs in the extended XY model
\cite{Li2018}. The QPTs and the quantum coherence in Heisenberg XYZ 
systems were not investigated carefully until now.
While most studies consider the ground-state factorization in symmetric
spin system, the knowledge is lacking in wondering the existence of
factorized ground states in more complex multipartite systems.
The primary motivation of the present work is to try to elucidate the
role of DM interaction in Heisenberg XYZ model, and explore whether
quantum criticality and factorization can be captured by emerging
coherence. Exploiting favorable figures of merit of quantum information
measures allows extracting the full ground-state phase diagram of the
spin-1/2 Heisenberg XYZ chain. We remark that especially two-qubit
reduced density matrices adopted in Ref. \cite{Rad17} are improper.

The remainder of this paper is organized as follows. We introduce the
1D anisotropic Heisenberg model with DM interactions in Sec.
\ref{sec:model}. In Sec. \ref{sec:mea}, we present the analytical
approach and calculate quantum entanglement and quantum coherence. 
In Sec. \ref{scaling}, we discuss the scaling behavior of the local
quantum coherence in the XY model, and the factorization phenomena
under the interplay of DM interactions and magnetic field.
Finally, in Sec. \ref{conclusion} we give the summary and conclusion.

\section{The model}
\label{sec:model}

We consider the anisotropic Heisenberg chain described by the
following Hamiltonian:
\begin{eqnarray}
    \cal{H}&=&  J\sum_{j=1}^{N}
    \left(\frac{1+\gamma}{2}\,{\sigma}^{x}_{j}{\sigma}^{x}_{j+1}+
    \frac{1-\gamma}{2}\,{\sigma}^{y}_{j}{\sigma}^{y}_{j+1}
    +\Delta {\sigma}^{z}_{j}{\sigma}^{z}_{j+1}\right)\nonumber\\
&+& \sum_{j=1}^{N}\vec{D} \cdot (\vec{\sigma}_{j}\times\vec{\sigma}_{j+1})
    -h\sum_{j=1}^{N}{\sigma}^z_j,,
\label{eq:ham}
\end{eqnarray}
where $N$ is the number of the spins in the chain, and the periodic
boundary condition is assumed, i.e., $\vec{\sigma}_{N+j}=\vec{\sigma}_j$,
and \mbox{$\vec{\sigma}_j=\{{\sigma}^x_j,{\sigma}^y_j,{\sigma}^z_j\}$}.
The model has AFM exchange coupling ($J\ge 0$), anisotropy $\Delta$,
DM vector $\vec{D}$, and uniform magnetic field strength $h$ acting on 
$\{\sigma^z_j\}$. Here we presume that the $\vec{D}$ vector is along 
the direction perpendicular to the plane, i.e., $\vec{D}=D \hat{z}$ and 
we take $D$ as a unit of $\vec{D}$. The parameter $\gamma\ge 0$
measures the anisotropy of spin-spin interactions in the $xy$ plane
which typically varies from 0 (isotropic XY model) to 1 (Ising model).

Several types of model interactions are currently being explored for
simulating effective spin systems like Ising, XY, and XYZ, which may
stand for systems of trapped ions \cite{Porras04} or polaritons
\cite{Berloff17}. An important extension of the effective models is
the DM interaction which can be interpreted as an electric field. 
The DM interaction was introduced by Dzyaloshinskii and Moriya in a 
phenomenological model~\cite{Dzy58} and a microscopic model 
\cite{Mor60}, respectively. The DM interactions exist in solids, such 
as ferrimagnetic insulator Cu$_2$OSeO$_3$ \cite{Seki12,Adams12,Yang12} 
or multiferroic BiFeO$_3$ \cite{Matsuda}, and are synthesized in 
optical lattices for both fermions~\cite{Wang12,Cheuk12} and bosons 
\cite{Garcia12,Cole12}.

A microscopic mechanism arises from that the electric polarization
generated by the displacement of oppositely charged ions is driven by
non-collinear spiral magnetic structures with a cycloidal component as
described by Tokura~\cite{Tokura10},
$\vec{P}\propto \hat{e}_{ij}\times(\vec{\sigma}_i\times\vec{\sigma}_j)$,
where $\hat{e}_{ij}$ is the unit vector connecting the neighboring
spins $\vec{\sigma}_i$ and $\vec{\sigma}_j$. The coupling coefficient
of macroscopic polarization is material-dependent \cite{Sergienko06},
and the sign depends on the vector spin chirality.
In this respect, an energy shift, $-\vec{D} \cdot \vec{P}$, by applying
an electric field $\vec{D}$ prevails over the Heisenberg exchange and
the QPT occurs in this system. The supplemented DM interaction can be
gauged away by performing a spin rotation with respect to a twist phase,
$\phi=\tan^{-1}(D/J)$ of spin operators,
$\sigma^+_j\sigma_{j+1}^-\to\sigma^+_j\sigma_{j+1}^- e^{i\phi}$, for
$\gamma=0$~\cite{Shekhtman92}. So, in this way the XXZ model has been
changed to the XYZ model after rotation. Note that the absence of
inversion symmetry in DM interaction introduces anisotropy to the
system.

\section{The information measures}
\label{sec:mea}

For general parameters, Hamiltonian (\ref{eq:ham}) is not integrable
except at specific points in parameter space. In the case of
$\gamma=0$, $D=0$, $h=0$, one finds that the XXZ spin chain with
nearest neighbor interaction is integrable, and it can be analytically
solved using the Bethe ansatz~\cite{Bethe1931,Yang1966}. Here we use
Jordan-Wigner transformation to represent the spin operators
$\sigma^{\pm}_{j}=(\sigma^x\pm i\sigma^y)/2$ by fermion
operators:
\begin{eqnarray}
\sigma _{j}^{+}& =&\exp\left[ i\pi\sum_{i=1}^{j-1}c_{i}^{\dagger }c_{i}^{}
\right] c_{j}^{}=\prod_{i=1}^{j-1}\sigma _{i}^{z}c_{j}^{},    \\
\sigma _{j}^{-}& =&\exp\left[-i\pi\sum_{i=1}^{j-1}c_{i}^{\dagger }c_{i}^{}
\right] c_{j}^{\dagger}=\prod_{i=1}^{j-1}\sigma_{i}^{z}c_j^{\dagger }, \\
\sigma _{j}^{z}& =&1-2c_{j}^{\dagger }c_{j}^{}.\label{eq:jw}
\end{eqnarray}

For $\Delta\neq 0$, we approximate the model (\ref{eq:ham}) by 
mean-field decoupling \cite{Vionnet2017}. In this approximation, 
the Ising coupling (four-fermion interaction) is
decomposed into three mean-field channels using Wick's theorem, which
are determined self-consistently in the noninteracting system:
\begin{eqnarray}
\label{eqMF}
 &&c^{+}_{j}c_{j}c^{+}_{j+1}c_{j+1}\nonumber\\ \nonumber &\approx&
\langle{c^{+}_{j}c_{j}}\rangle	c^{+}_{j+1}c_{j+1}
+ \langle{c^{+}_{j+1}c_{j+1}}\rangle c^{+}_{j}c_{j}
- \langle{c^{+}_{j}c_{j}}\rangle \langle{c^{+}_{j+1}c_{j+1}}\rangle 	
 \nonumber \\
&-& \langle{c^{+}_{j}c^{+}_{j+1}}\rangle c_{j}c_{j+1}
- \langle{c_{j}c_{j+1}}\rangle c^{+}_{j}c^{+}_{j+1}
+ \langle{c^{+}_{j}c^{+}_{j+1}}\rangle\langle{c_{j}c_{j+1}}\rangle
 \nonumber \\
&+&\langle{c^{+}_{j}c_{j+1}}\rangle c_{j}c^{+}_{j+1}
+ \langle{c_{j}c^{+}_{j+1}}\rangle c^{+}_{j}c_{j+1}
- \langle{c^{+}_{j}c_{j+1}}\rangle \langle{c_{j}c^{+}_{j+1}}\rangle.
 \nonumber \\
\end{eqnarray}
We mainly use the mean-field approximation of the translation invariant
Hamiltonian to get the three self-consistent parameters $\{\mu,t,\delta\}$
(see below),
\begin{eqnarray}
\cal{H}&=&\sum_{j=1}^{N}\left\{
\left[J c_{j+1}^+ c_j+J\gamma c_{j+1}^+ c^+_j+{\rm h.c.}\right]\right. \nonumber \\
&&\left.+J \Delta (1-2c_j^+c_j)(1-2c_{j+1}^+c_{j+1})-h(1-2c_j^+c_j)\right\}\nonumber \\
&\approx&\sum_{j=1}^{N} \left\{
[(t+2iD)c_{j+1}^+c_j+\delta c^+_jc_{j+1}^++{\rm h.c.}]+\mu c_j^+c_j\right\}\nonumber \\
&&+{\rm const},
\label{eq:ham3}
\end{eqnarray}
i.e., $\mu=4J\Delta (2\langle{c^+_{j}c_{j}}\rangle-1)+2h$,
$t=J(1-4 \Delta\langle{c^+_{j}c_{j+1}}\rangle)$, and
$\delta=J(\gamma-4\Delta\langle{c_{j}c_{j+1}}\rangle)$. The values of
$\{\mu,t,\delta\}$ can be determined self-consistently
(see Appendix \ref{appendix-MF}).

First, we discuss the model Eq. (\ref{eq:ham}) at $\Delta=0$, where it
reduces to the XY chain. The XY model is an archetypal model of quantum
physics, encapsulating the physics underlying universal phenomena in
equilibrium phase transition, much less looking for exotic phases
through machine learning
\cite{juan1,un1,un2,Sca17,zhai17,beach18,zhai18,Zha18}.
On one hand, the linear XY chain in the presence of transverse fields
plays a central role in condensed matter theory
\cite{Rad18,Toskovic16} and is a good candidate for building blocks in
quantum computers~\cite{Zheng2000,Christandl04} and quantum information
applications~\cite{Bose2003}. On the other hand, and maybe even more
importantly, the XY model is one of few exactly solvable models in
strongly correlated systems, and thus provides the benchmark for other
approximate techniques implemented in more realistic models, especially
for the accurate calculation of various dynamic quantities.

While being relatively simple, the XY model exhibits a rich phase
diagram. Applying the transverse field induces an Ising transition
at $h_c=1$ from AFM phase to paramagnetic (PM) phase.
Moreover, a completely factorized ground state may occur at a specific
value of the field,
\begin{equation}
h_f=J \sqrt{\left(\frac{1+\gamma}{2}+\Delta\right)
            \left(\frac{1-\gamma}{2}+\Delta\right)},
\end{equation}
in the absence of DM interaction \cite{Giampaolo08}.

\begin{figure}[t!]
  \includegraphics[width=.99\columnwidth]{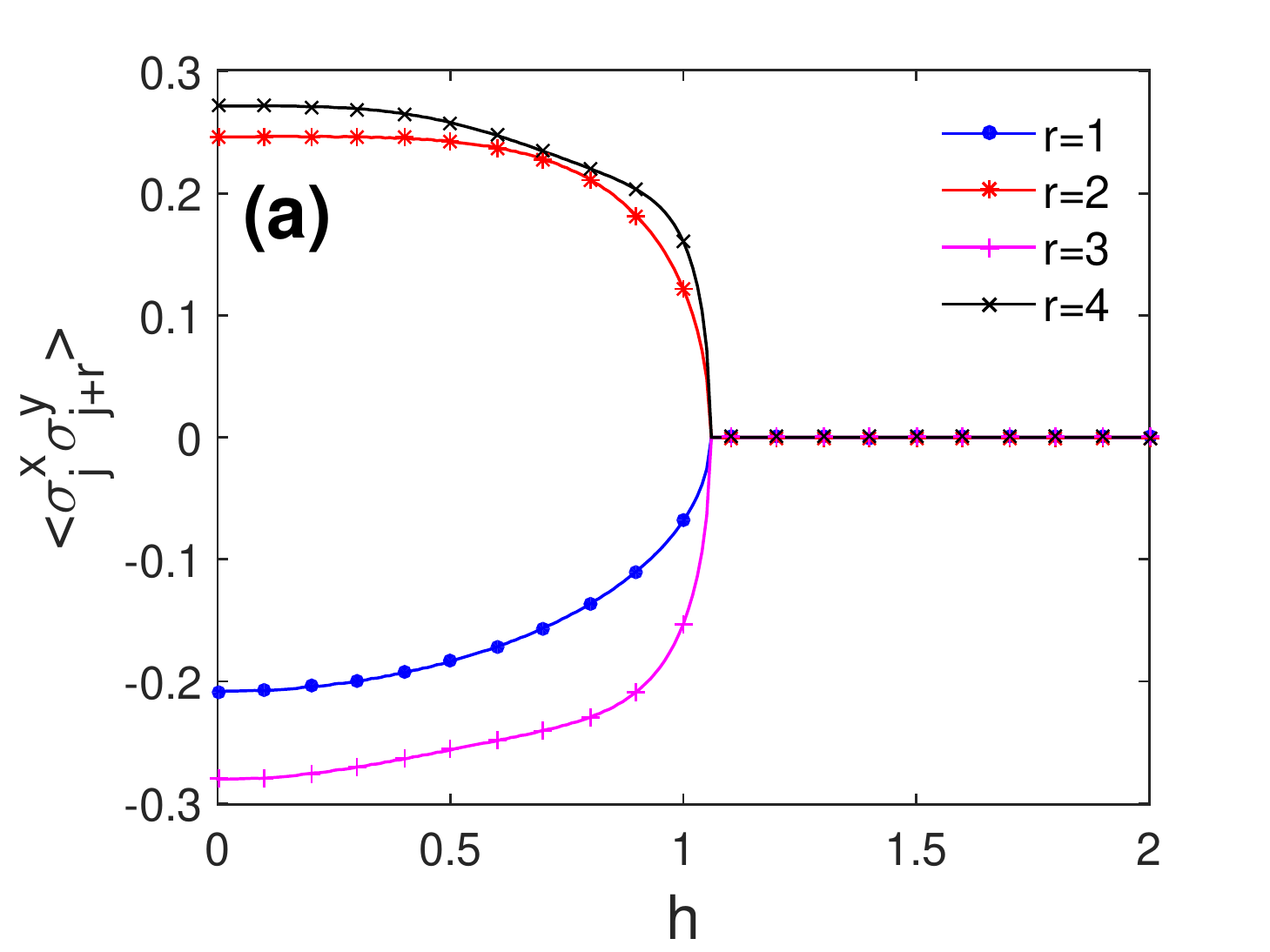}
  \includegraphics[width=.99\columnwidth]{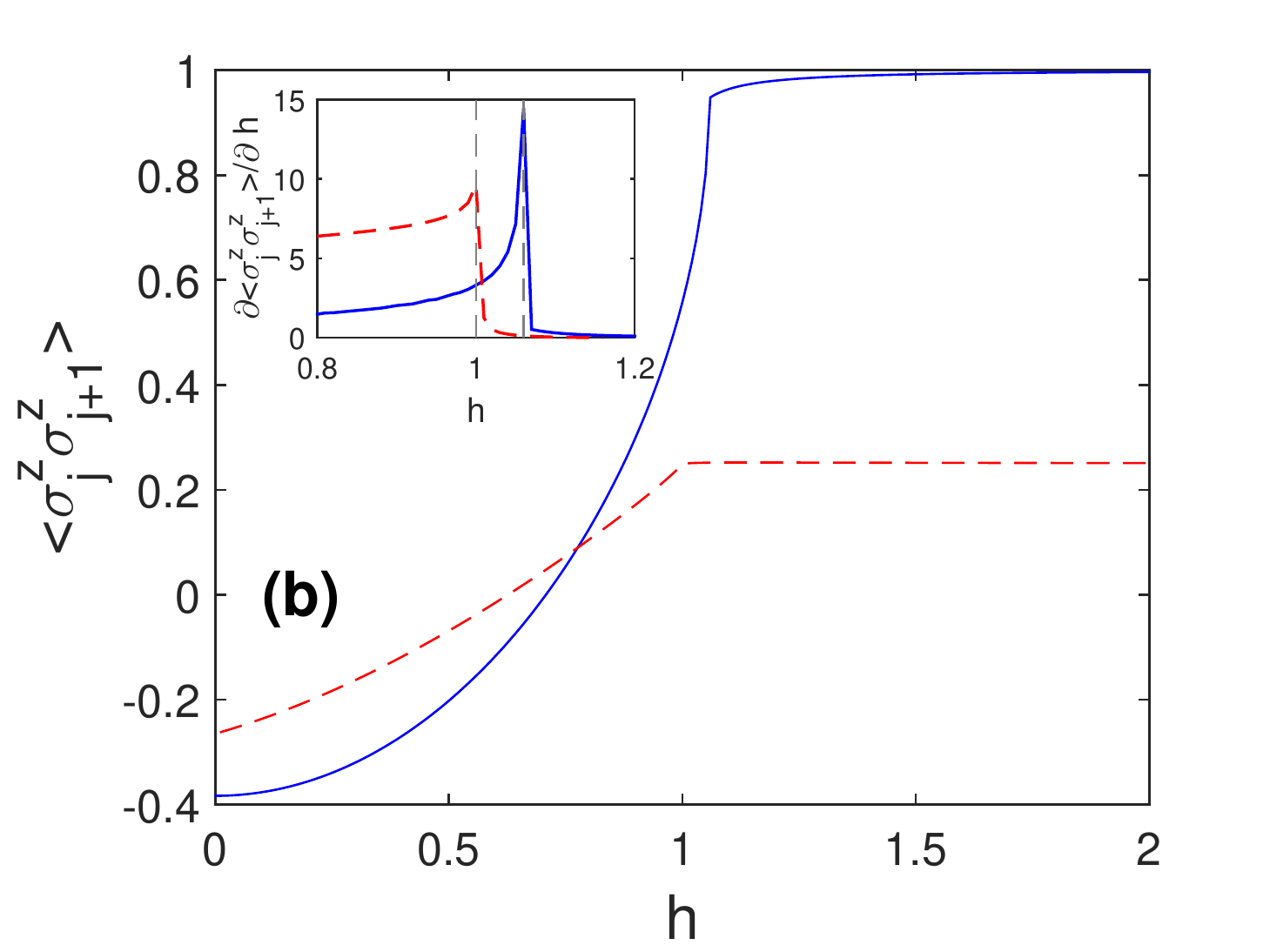}
  \includegraphics[width=.99\columnwidth]{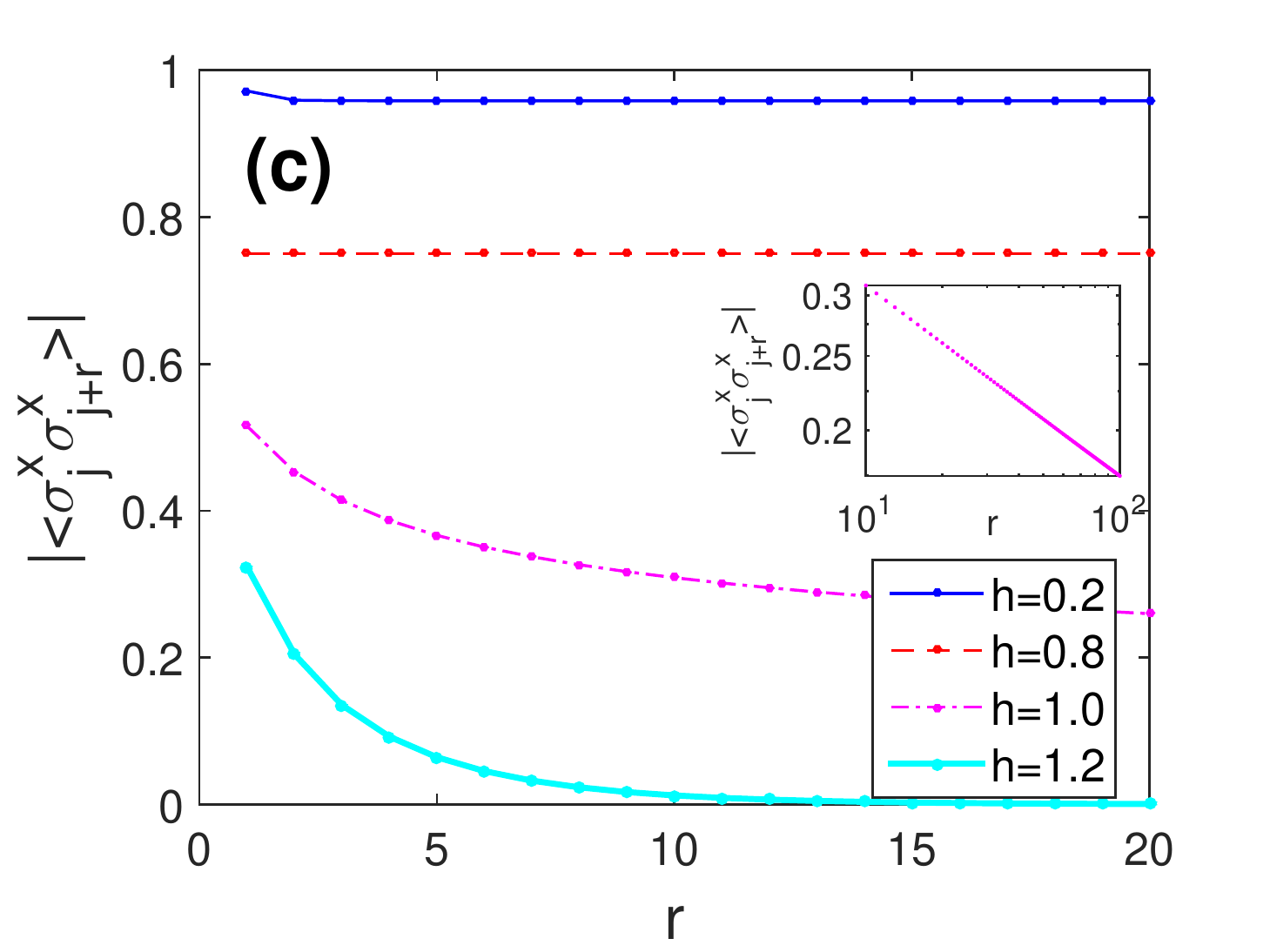}
\caption{Short-range correlations for increasing field $h$:
(a)~two-qubit correlations $\langle\sigma_j^x\sigma_{j+r}^y\rangle$
for different distance with $\gamma=0.2$, and
(b)~the~nearest neighbor correlation,
$\langle\sigma_j^z\sigma_{j+1}^z\rangle$ with $\gamma=0.2$.
The solid line is plotted according to Eq. (\ref{eq:zzcorr}),
while the dashed line is obtained by assuming
$\langle\sigma_j^x\sigma _{j+1}^y\rangle=0$.
Inset shows the corresponding first derivative.
(c) The absolute value of the correlation function,
$|\langle\sigma_j^x\sigma_{j+r}^x\rangle|$ with $\gamma=0.6$, 
for increasing~$r$. Other parameters: $D=0.2$, $\Delta=0$, $J=1$. }
\label{fig:Corr}
\end{figure}

We are now in a position to explore the quantum coherence and
entanglement measures based on the reduced density matrix. They can be
determined without a full tomography of the state under consideration.
The studies of quantum entanglement and coherence are crucial for both
fundamental issues and niche technological applications. We consider
the representation spanned by the two-qubit product states:
$\vert 1\rangle\equiv\vert{\uparrow}\rangle_i\otimes\vert{\uparrow}\rangle_j$,
$\vert 2\rangle\equiv\vert{\uparrow}\rangle_i\otimes\vert{\downarrow}\rangle_j$,
$\vert 3\rangle\equiv\vert{\downarrow}\rangle_i\otimes\vert{\uparrow}\rangle_j$,
and
$\vert 4\rangle\equiv\vert{\downarrow}\rangle_i\otimes\vert{\downarrow}\rangle_j$.
Here $\vert{\uparrow}\rangle$ ($\vert{\downarrow}\rangle$) stands for
spin up (down) state, and the reduced density matrix for selected two
auxiliary qubits can be expressed in the following form:
\begin{equation}
\rho_{ij}=\left(
\begin{array}{cccc}
u^{+} & 0 & 0 & z_{1} \\
0 & w_{1} & z_{2} & 0 \\
0 & z_{2}^{*} & w_{2} & 0 \\
z_{1}^{*} & 0 & 0 & u^{-}
\end{array}
\right),  \label{eq:2DXXZ_RDM}
\end{equation}
with
\begin{eqnarray}
u^{\pm }\!&=&\frac{1}{4}\left(1\pm  2\langle { \sigma_{i}^{z}}\rangle
+\langle { \sigma _{i}^{z} \sigma _{j}^{z}}\rangle \right),
\label{upm} \\
z_{1}\!&=&\frac{1}{4}\left(
  \langle\sigma_{i}^{x}\sigma_{j}^{x}\rangle
- \langle\sigma_{i}^{y}\sigma_{j}^{y}\rangle
-i\langle\sigma_{i}^{x}\sigma_{j}^{y}\rangle
-i\langle\sigma_{i}^{y}\sigma_{j}^{x}\rangle\right),  \label{z1} \\
z_{2}\!&=&\frac{1}{4}\left(
  \langle\sigma_{i}^{x}\sigma_{j}^{x}\rangle
+ \langle\sigma_{i}^{y}\sigma_{j}^{y}\rangle
+i\langle\sigma_{i}^{x}\sigma_{j}^{y}\rangle
-i\langle\sigma_{i}^{y}\sigma_{j}^{x}\rangle\right),  \label{z2} \\
\omega_{1}\!&=&\omega_{2}=\frac{1}{4}\left(
1-\langle\sigma_{i}^{z}\sigma_{j}^{z}\rangle\right).  \label{omega1}
\end{eqnarray}
A representative state $X$ stands for a five-parameter family of states
of two qubits. Be aware that the reflection symmetry is broken in
Hamiltonian (\ref{eq:ham}) thanks to the existence of DM terms. In this
case,
$\langle\sigma_{i}^{x}\sigma_{j}^{y}\rangle$ and
$\langle\sigma_{i}^{y}\sigma_{j}^{x}\rangle$ in Eqs. (\ref{z1}) and
(\ref{z2}) are not necessarily vanishing, see Fig.~\ref{fig:Corr}(a).
Nevertheless, a simplification that
$\langle\sigma_{i}^{x}\sigma_{j}^{y}\rangle=0$ and
$\langle\sigma_{i}^{y}\sigma_{j}^{x}\rangle=0$ was commonly used in
the past \cite{Li2009,Liu11,Rad17,Ye2018}.
Also, such negligence in calculating Eqs. (\ref{upm}) and (\ref{omega1})
frequently occur in terms of the relation \cite{Li2009,Liu11,Rad17,Ye2018}:
\begin{eqnarray}
\langle\sigma_{j}^{z}\sigma_{j+1}^{z}\rangle&=&
\langle\sigma_{j}^{z}\rangle\langle\sigma_{j+1}^{z}\rangle
-\langle\sigma_{j}^{x}\sigma_{j+1}^{x}\rangle
\langle\sigma_{j}^{y}\sigma_{j+1}^{y}\rangle
\nonumber \\
&+&\langle\sigma_{j}^{x}\sigma_{j+1}^{y}\rangle
\langle\sigma_{j}^{y}\sigma_{j+1}^{x}\rangle.
\label{eq:zzcorr}
\end{eqnarray}
As is shown in Fig. \ref{fig:Corr}(b), the inclusion of the last term
in Eq. (\ref{eq:zzcorr}) brings a prominent difference. For instance,
the first-derivative of nearest neighbor correlation accurately
discriminates the criticality. A two-qubit state tomography can be
implemented on the system qubits~\cite{Liu2016}, and holds advantage
that a full tomography of the state is not necessary.

Quantum coherence is a kind of quantification of quantum superposition,
which is one of the most significant properties of quantum states
separate from classical ones. As the core of quantum physics and
quantum information, there are many important applications in various
quantum tasks, such as quantum computation and quantum communication.
There are many related studies
\cite{Liu11,Rad17,You16,You17,Lei2015}. A well-defined and frequently
used coherence measure is Wigner-Yanase skew information (WYSI),
which has some clear physical meanings, such as it is equal to the
optimal distillation rate for standard coherence distillation, and
can also be interpreted as the minimal amount of noise required to
achieve full decoherence of the state under discussion. WYSI mainly
quantifies the information encapsulated in a quantum state with
respect to an observable $K$ \cite{Winger1963,Gir14}, which has
implemental value in both theoretical and
experimental schemes in view of the current technology:
\begin{eqnarray}
{\cal I}(\rho,K)=-\frac{1}{2} \textrm{Tr}[\sqrt{\rho},K]^2,
\label{WYSI}
\end{eqnarray}
where [.,.] stands for the commutator.
The WYSI was interpreted as a measure quantifying the non-commutativity
between $\rho$ and $K$ \cite{Luo2012}, and thus captures the genuine
quantum uncertainty of a given observable in a certain quantum state.
Very recently it has been proven by Girolami \cite{Gir14} that
${\cal I}(\rho,K)$ given by Eq. (\ref{WYSI}) satisfies all the criteria
for coherence monotones \cite{Baumgratzq14} and consequently can be 
used as a reliable measure of coherence.

We find that the QPT and
factorization phenomenon are both associated with the local quantum
coherence (LQC)~\cite{Karpat2014,Li2018}, as quantified by WYSI,
in single-spin and two-spin reduced density matrices of the ground
state of the spin chain. For a bipartite system, the LQC describes the
observable that acts only on one of the subsystems, as
${\cal I}(\rho_{AB},K_{A}\otimes I_{B})$. Here we choose $K_{A}$ as
$\sigma^x$ or $\sigma^z$. Karpat, Cakmak, and Franchini found that the
WYSI remains non-increasing under classical mixing of quantum states
\cite{Karpat2014}. It filters out the pure
quantum uncertainty in a measurement.

The absence of the WYSI implies
that no quantum uncertainty can be observed, and statistical errors
are due to classical ignorance. Analogously, the concurrence is a
pairwise entanglement measure for any bipartite system that relates to
the two-site reduced density matrix $\rho$~\cite{Wootters}.
The concurrence for a two-qubit state $\rho_{ij}$ is defined as
$C$=2 max$\{$0, $\Lambda_1$,$\Lambda_2 \}$, where
$\Lambda_1=\vert z_1\vert-\sqrt{\omega_1\omega_2}$ and
$\Lambda_2=\vert z_2\vert-\sqrt{u^{+}u^{-}}$~\cite{Werlang09}.

\section{RESULTS}
\label{scaling}

\begin{figure}[t!]
\noindent
\includegraphics[width=1.05\columnwidth]{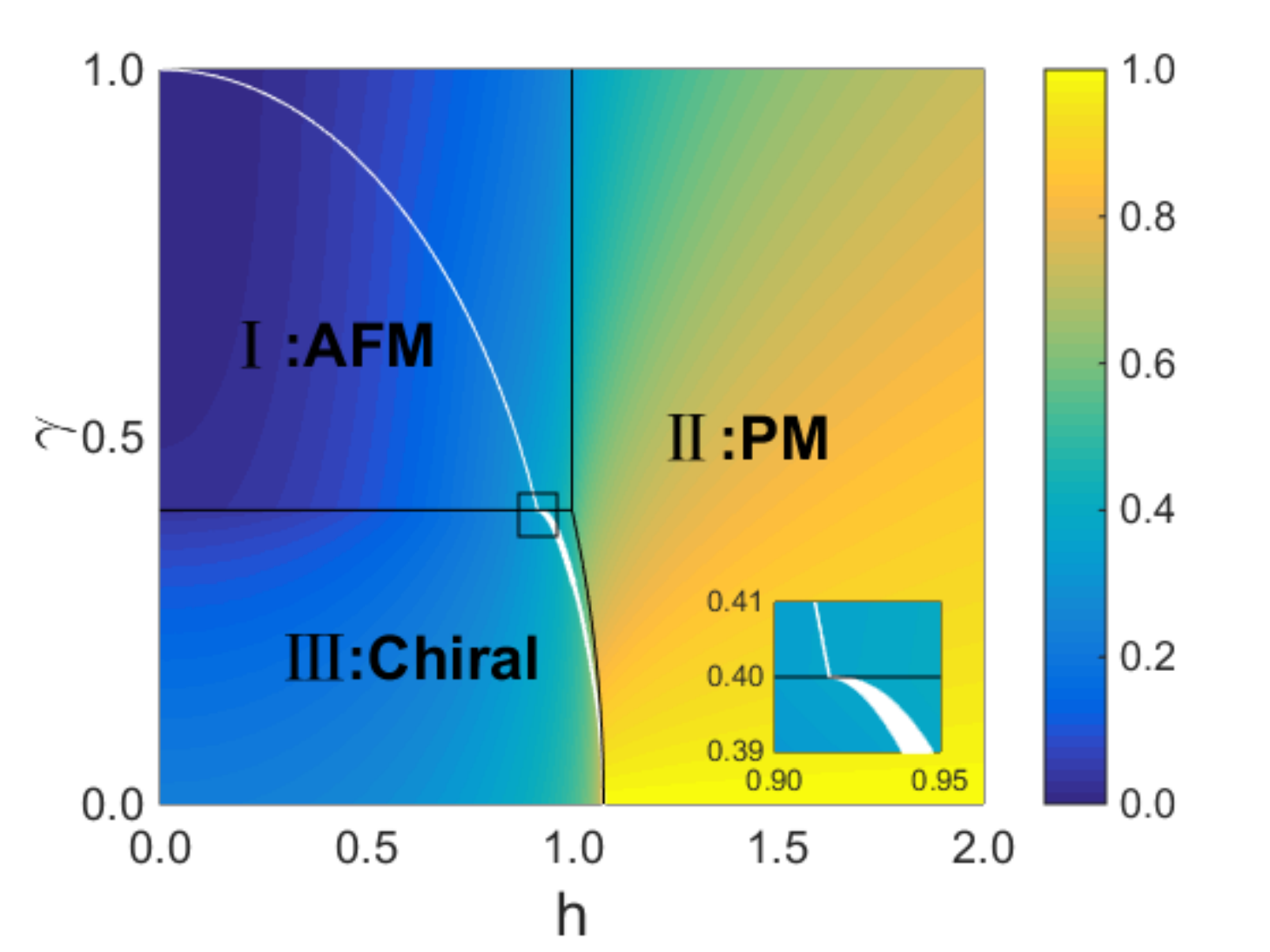}
\caption{Magnetic phase diagram of the 1D XY model with three
(black solid) lines separating the phases I, II, and III.
In the special case of $D=0$, the chiral phase vanishes.
The color map represents the strength of the LQC$_x$,
${\cal I}(\rho_{j,j+1},\sigma_j^x)$. The white zone in the AFM phase
and the chiral phase delegates the factorization line and the
factorization region, respectively.
Parameter: $D=0.2$, $\Delta=0$, and $J=1$.}
\label{fig:PD}
\end{figure}

To understand better the ground state in different regimes of
parameters it is natural to utilize diverse quantum information measures
to unfold the landscape of the criticality and factorization under the
effect of DM interactions and $\Delta$. In order to fully appreciate
the diversity of solutions, it is sufficient to study the special cases
with the spin-spin interactions in the $xy$ plane. For $\Delta=0$,
the diagonalization procedure of the left flip-flop couplings can be
achieved by the well-established techniques including Jordan-Wigner,
Fourier, and Bogoliubov transformations (see Appendix
\ref{appendix-Exactsolution}). Using the exact solutions, the
correlation functions could be obtained, and the magnetic phase
diagram presented in Fig. \ref{fig:PD} was found.

The phase diagram consists of three phases:
AFM phase I, PM phase II, and gapless Chiral phase III.
The phase boundaries are determined by three lines: $h=1$, $\gamma=2D$,
and $h=\sqrt{4D^2-\gamma^2+1}$, respectively~\cite{Yi2018}.
As shown in Fig.~\ref{fig:Corr}(c), in the AFM phase, the correlation
function $\langle\sigma_i^x\sigma_{i+r}^x\rangle$
becomes a constant quickly, although there is a small decrease
for $r\ge 2$ comparing with the nearest neighbor correlation. At the
critical point, i.e., at $h=1.0$, the correlation function has an
algebraic decay, namely,
$\langle\sigma_i^x\sigma_{i+r}^x\rangle\sim r^{-1/4}$.
On the contrary, the correlation function decays exponentially with 
the increase of $r$ in the PM phase.

\begin{figure}[t!]
\noindent
 \includegraphics[width=\columnwidth]{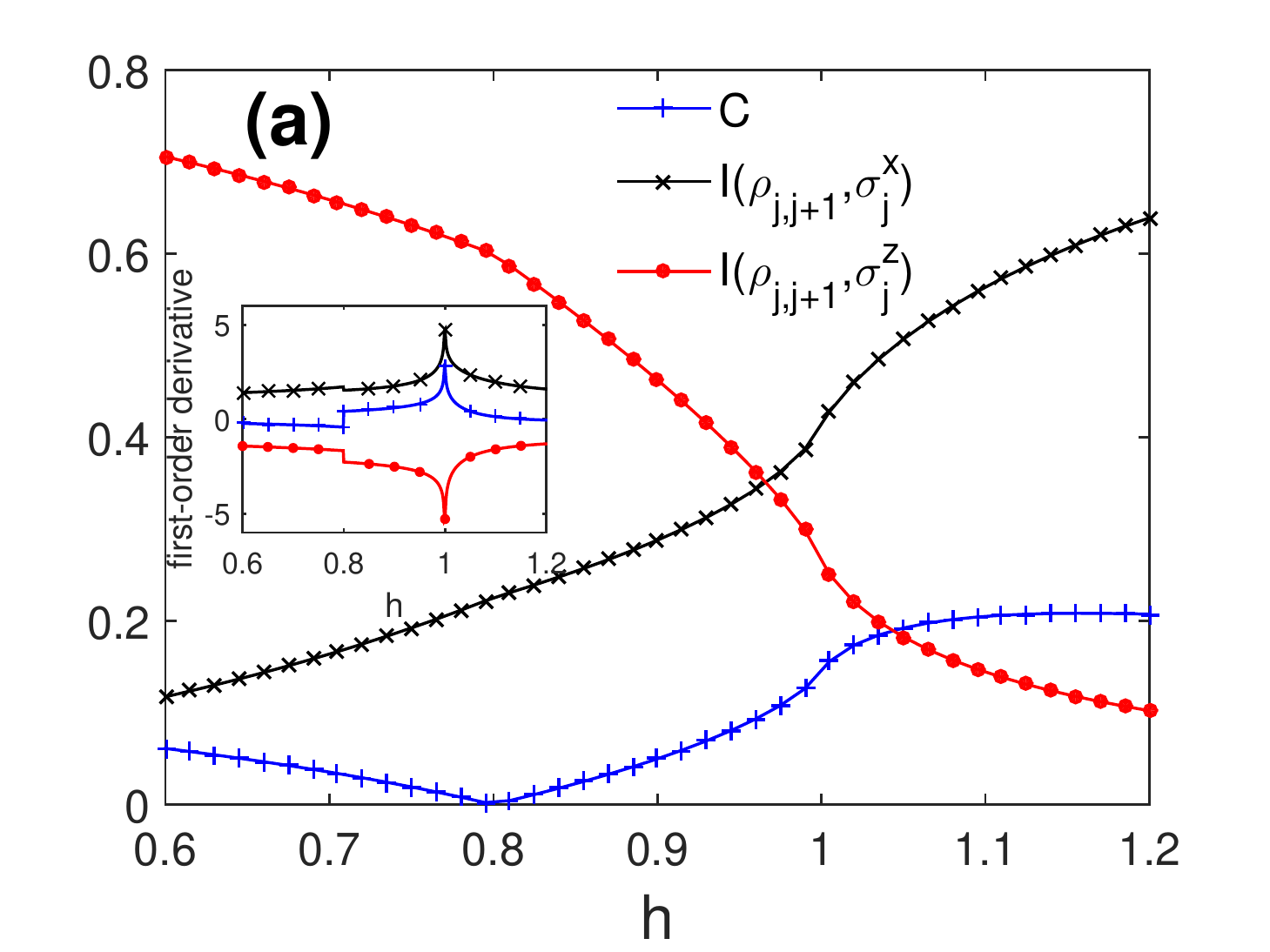}
 \includegraphics[width=\columnwidth]{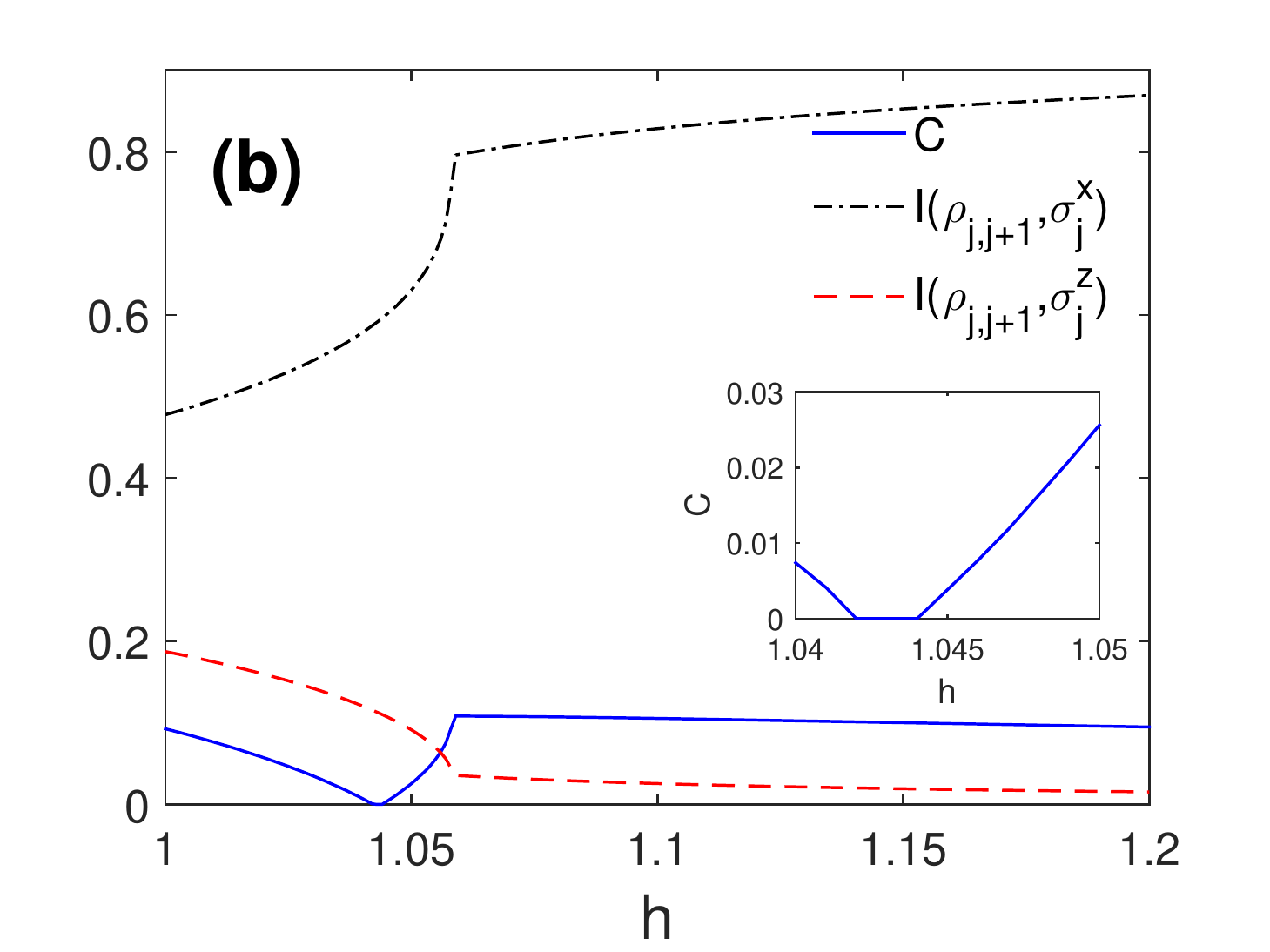}
\caption{The concurrence, LQC$_x$ and LQC$_z$ with respect to $h$ at
$D=0.2$, and for:
(a) $\gamma$=0.6;
(b) $\gamma$=0.2.
Inset in (a) shows the first-derivative of the concurrence, LQC$_x$ and
LQC$_z$.
Inset in (b) shows the magnified factorized zone.
Parameter:  $\Delta=0$, $J=1$.}
\label{fig:entanglement}
\end{figure}

Figure \ref{fig:entanglement} examines the local quantum $\sigma^x$
coherence (LQC$_x$) and the local quantum $\sigma^z$ coherence (LQC$_z$)
along $\gamma=0.6$ and $\gamma=0.2$, respectively, with $D=0.2$, which
corresponds to AFM-PM and chiral-PM transitions. The LQC$_x$ is
monotonously increasing with $h$, in contrast to the monotonous decay
of LQC$_z$. The LQC$_x$ is indeed large for large $h$, especially for
$\gamma=0$, where LQC$_x$ dramatically increases to unity when $h$
approaches $h_c$. Instead, the concurrence shows the non-monotonous
characteristics of entanglement. As $h$ increases, the concurrence
firstly decreases to zero and then increases with $h$.  Although the
entanglement and coherence measures do not exhibit any divergences, the
divergences of their first derivatives with respect to $h$ may be used
to identify the critical points. Indeed, the derivative of the quantum
information measures has been proven to be
a powerful tool to detect the location of the quantum critical points
\cite{Woo82,Ollivier2001,Ost02,Osborne2002,Gu2003,Vidal2003,Fan18}.

In the inset of Fig. \ref{fig:entanglement}(a), one observes that the
first derivatives of both LQC$_x$ and LQC$_z$ show a cusp singularity
at $h_m$, which marks the point of the QPT. It is even more evident in
Fig. \ref{fig:entanglement}(b) that the critical points can be identified
by kink behavior in these information measures. In this case, their
first derivatives are discontinuous. Note that the quantum
Jensen-Shannon divergence was adopted to inspect the quantum coherence
of the 1D XY model with DM interactions~\cite{Rad17}. However, an
incomplete phase diagram is identified due to the illegal results of
$\langle\sigma_i^x\sigma_j^y\rangle$,
$\langle\sigma_i^y\sigma_j^x\rangle$,
$\langle\sigma_i^z\sigma_j^z\rangle$
in the presence of DM interactions, as is clearly demonstrated in Fig.
\ref{fig:Corr}(a-b). We remark that $\langle\sigma_i^x\sigma_j^y\rangle=0$
and $\langle\sigma_i^y\sigma_j^x\rangle=0$ should not be taken for
granted in general for a system with broken reflection symmetry.

\begin{figure}[t!]
\noindent
 \includegraphics[width=\columnwidth]{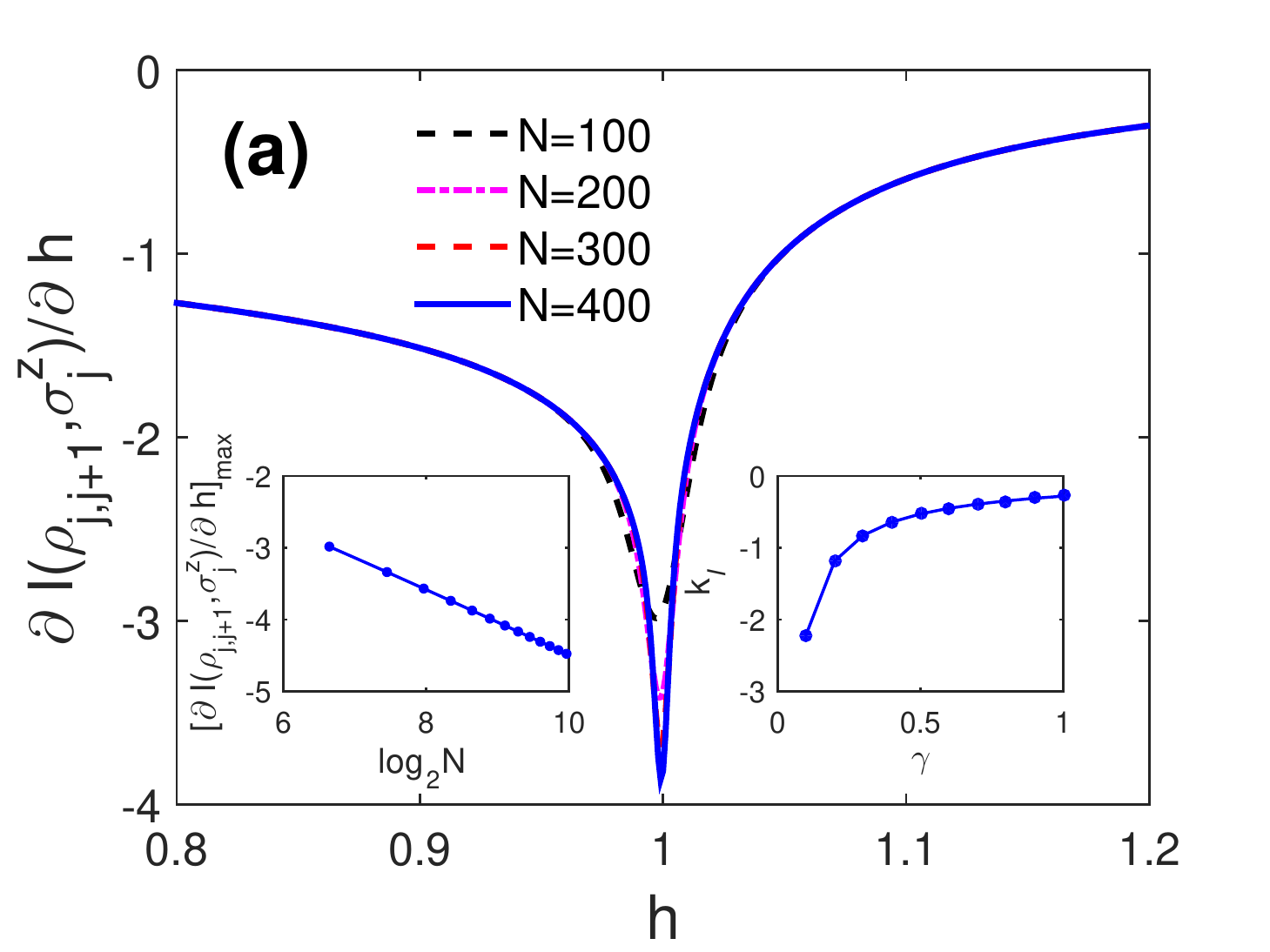}
 \includegraphics[width=\columnwidth]{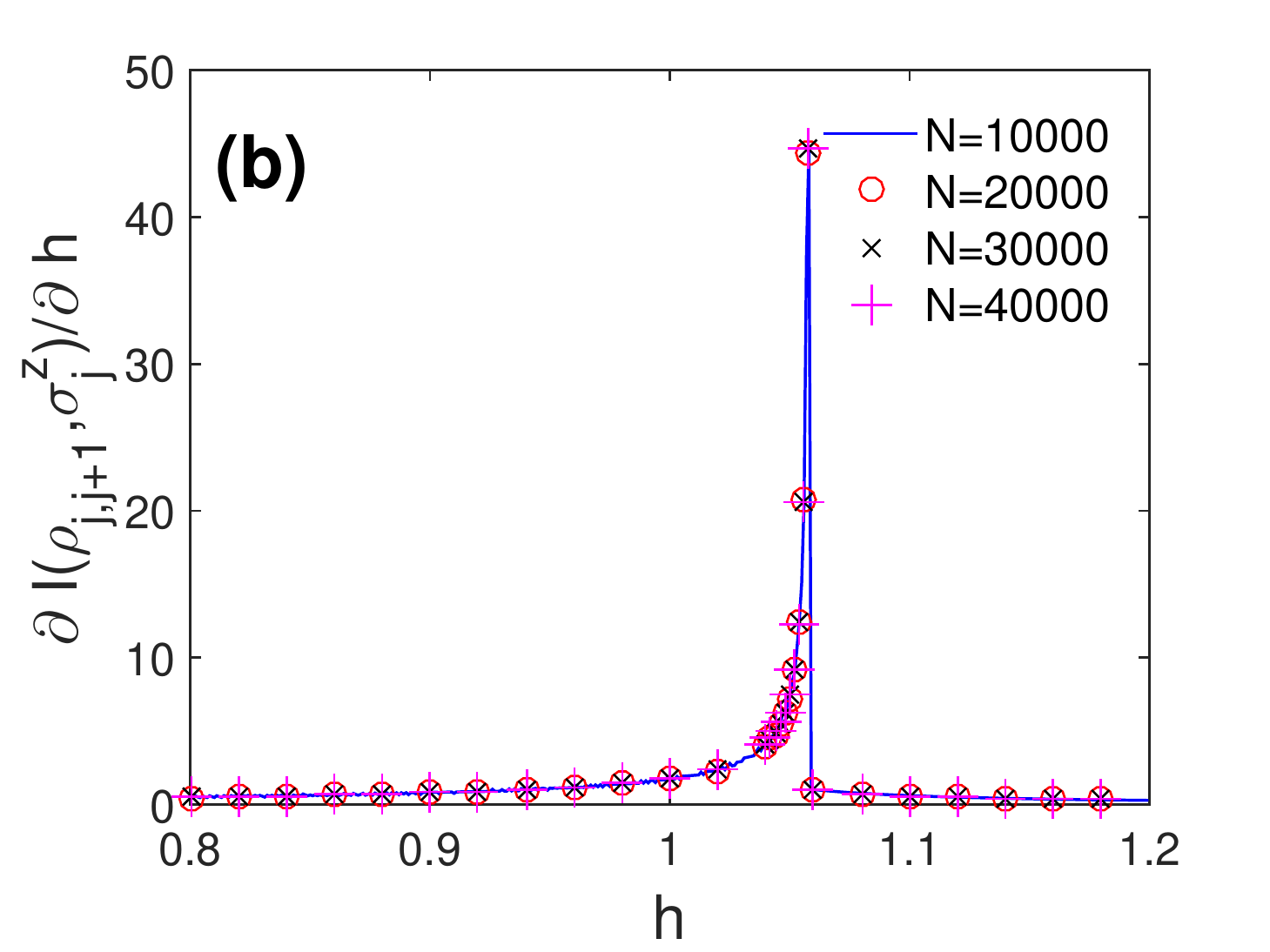}
\caption{The first derivative of LQC$_z$ with respect to $h$ for
different system sizes with:
(a) $\gamma$=0.6, and
(b) $\gamma$=0.2. In the left inset of (a), the maxima of the peaks 
(dots) follow a logarithmic scaling $\left(\frac{\partial{\cal I}}
{\partial h}\right)_{\text{max}}=0.382\log_2{N}-0.056$ (solid line). 
The right inset shows the dependence of $k_1$ in Eq.
(\ref{definek1}) on $\gamma$.  
Other parameters: $D=0.2$, $\Delta=0$, and $J=1$.
}
\label{fig:scaling}
\end{figure}

Along with the location of quantum critical points, the critical
exponents can also be extracted by the scaling of quantum information
measures. The first-order derivatives of the two-spin local $\sigma^z$
coherence with respect to $h$ are shown in Fig.~\ref{fig:scaling}(a).
We notice that the first-order derivative around the critical point
becomes sharper and sharper as the system size increases, and it is
expected to diverge in the thermodynamic limit.
The first-order derivative of the LQC$_z$ follows a logarithmic
divergence across the critical point,
\begin{eqnarray}
\label{definek1}
{\left(\frac{\partial{\cal I}}{\partial h}\right)}_{\rm max}
\propto k_1\log_{2}{N},
\end{eqnarray}
as is disclosed in the inset of Fig. \ref{fig:scaling}(a).
Here $k_1$ is a constant and is monotonically decreasing with $\gamma$.

We analyzed the relative entropy, the concurrence and the logarithmic
negativity, and find their first-order derivatives obey similar
logarithmic scalings with different $k_1$, which are consistent with
the results in Refs. \cite{Ost02,Rad18}. However, for the QPT between
gapless chiral phase and gapped PM phase, the derivatives of the LQC
have pronounced peaks [see Fig. \ref{fig:scaling}(b)] which appear
independent of system size. The location of the pseudocritical field
$h_m$ approaches the true critical point $h_c$ as $N\to\infty$. Due to
the relevance of the driving Hamiltonian under the renormalization
group transformation, the leading term in the expansion of
pseudocritical point for sufficiently large systems obeys such scaling
behavior as in Refs. \cite{Ost02,Rad18},
\begin{eqnarray}
|h_m-h_c|\sim N^{-\alpha}.
\end{eqnarray}
By applying linear regression to the raw data obtained from
${\cal I}(\rho_{i,i+1},\sigma_i^z)$ on various system sizes,
one obtains \mbox{$\alpha=1.6642$} for $\gamma=1$.

Alternately, slightly away from the critical point in the thermodynamic
limit, the first field derivative of ${\cal I}$ satisfies,
\begin{eqnarray}
{\left(\frac{\partial {\cal I}}{\partial h}\right)}
\sim k_2 \log_2 \vert h-h_c\vert.
\end{eqnarray}
According to the scaling \textit{Ansatz} for logarithmic scaling
\cite{Barber83}, the ratio of $\vert k_1/k_2\vert$ gives rise to the
correlation length exponent $\nu$. Similar results were obtained as
well by the earlier studies \cite{Ost02,Zhu06,Rad18}.
The values resulting from different measures are consistent with each
other up to two digits~\cite{Tong16}. The results for $\gamma=0.6$ and
1.0, with $D=0.2$, are listed in Table \ref{table1}, which suggests
$\nu\simeq 1$ for Ising transition.

\begin{table}[b!]
\caption{Fitting parameters $\{k_1,k_2,\nu\}$ of the slope in logarithmic
scaling across the critical points with $D=0.2$, $\Delta=0$, and $J=1$. 
\label{table1}
}
\begin{ruledtabular}
\begin{tabular}{cccc}
$\gamma$ & parameter & $C$ & ${\cal I}(\rho_{j,j+1},\sigma^z_j)$ \\
\hline
 0.6 & $k_1$ & $\quad 0.33813$ & 0.45847 \\
     & $k_2$ & $\,   -0.33893$ & 0.44162 \\
     & $\nu$ & $\quad 0.99764$ & 1.03815 \\
\hline
 1.0 & $k_1$ & $\quad 0.18655$ & 0.28532 \\
     & $k_2$ & $\,   -0.18548$ & 0.30575 \\
     & $\nu$ & $\quad 0.99426$ & 1.07160 \\
\end{tabular}
\end{ruledtabular}
\end{table}

A close inspection of Fig.~\ref{fig:entanglement}(a) reveals there is
an interesting phenomenon simultaneously at $h_f=0.8$, where
the concurrence becomes zero and the LQC has a jump discontinuity.
The null point of the concurrence implies the ground state is
disentangled at this point, where the ground state simplifies into 
simple product states, i.e.,
$|\psi_0\rangle=\prod_{i=1}^{{N}}\otimes|\psi_{i}\rangle$, where
$|\psi_{i}\rangle$ are the states of the spins on the $i$th site.
Such product states lie exactly on a classical line $\gamma^2+h^2=1$
in the absence of the interaction $\propto D$
\cite{Kur82,Campbell13,Wei2005}, where the intersite correlations are
independent of the distance $r$ of two qubits, see Fig.
\ref{fig:Corr}(c).

\begin{figure}[t!]
\noindent
\includegraphics[width=1.05\columnwidth]{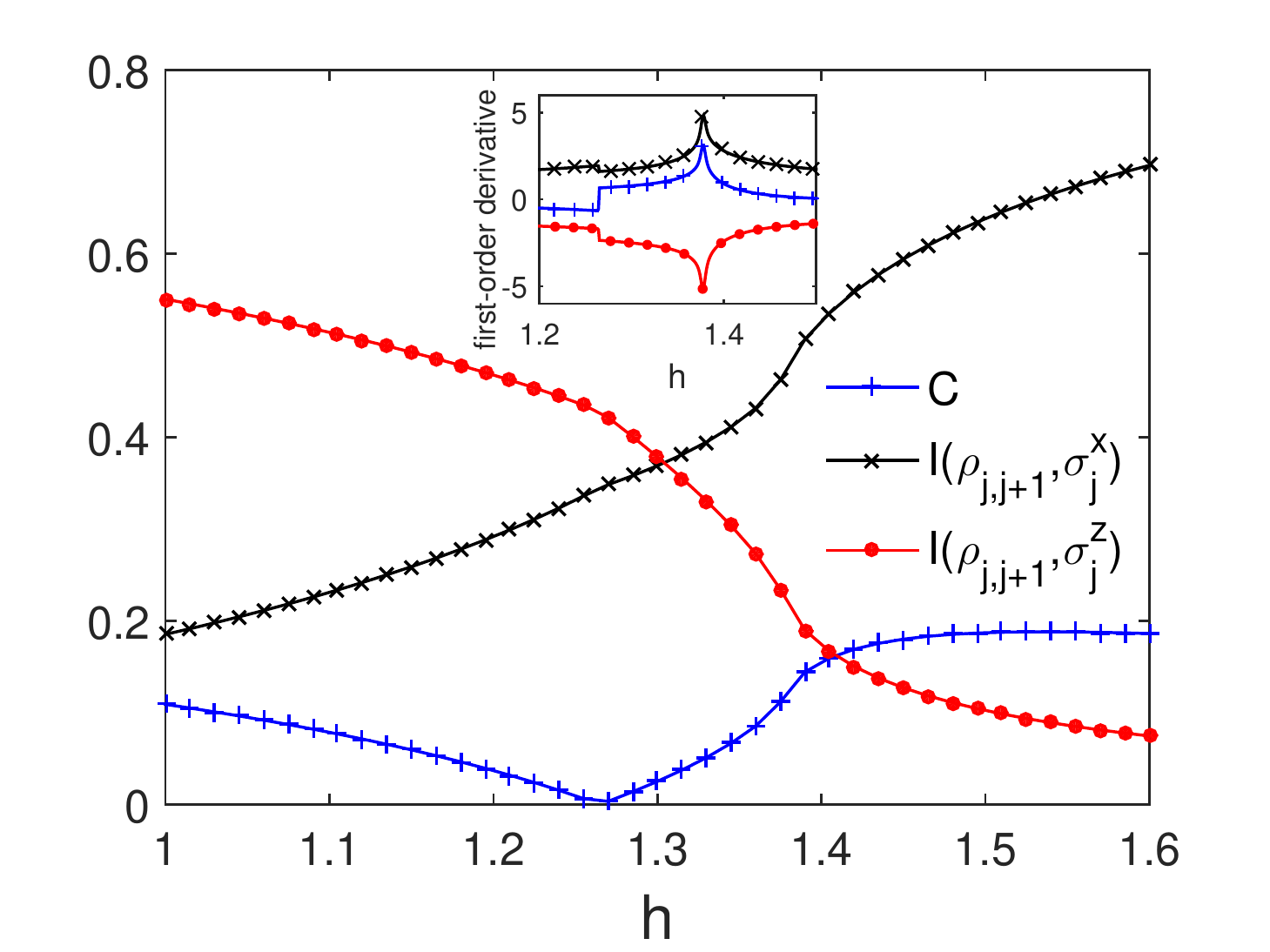}
\caption{The concurrence, LQC$_x$ and LQC$_z$ with respect to $h$.
Inset shows their first-derivative.
Parameters: $\gamma=0.6$, $D=0.2$, $\Delta=0.2$, and $J=1$. }
\label{fig:Heisenberg_coherence}
\end{figure}

As such, we find the logarithmic negativity is also able to mark the
classical feature while the von Neumann entropy fails to spot
vanishing entanglement. Notice that ground states of interacting spin
systems in the presence of an external magnetic field are typically
entangled and a completely separable ground state emerges only under
strict conditions. The exceptional phenomenon of separable ground state
has been thoroughly investigated in spin systems immersed in a uniform
transverse field~\cite{Muller1985,Roscilde05,Giampaolo08},
and nonuniform field~\cite{Cerezo16,Cerezo17}.
It was recognized that the factorization is a consequence of 
ground-state parity transition. Moreover, the behavior of the 
correlation functions changes from monotonic decay to oscillatory tail 
across the factorization point~\cite{Bar70}. On this line,
by examining the ground state of finite-size systems, the coherence
and entanglement witness remain constant for all values of $r$ at $h_f$
in the thermodynamic limit, when the system is actually in
the N\'{e}el phase \cite{Campbell13}.

In the chiral phase, shown in Fig. \ref{fig:entanglement}(b),
the concurrence also exhibits a similar trend with the increase of $h$.
It is odd to find that in this case the concurrence is vanishing for a
finite range of $h$. In addition, the range gets narrower as $\gamma$
decreases. The factorization volume in chiral phase is connected with
the factorization line in the AFM phase across the critical boundary
$\gamma=2D$, as displayed in the inset of Fig. \ref{fig:PD}. 
Also, the correlation functions are not constant anymore, and instead, 
the amplitude of correlation functions, including
$\langle\sigma_{i}^{x}\sigma_{i+r}^{x}\rangle$,
$\langle\sigma_{i}^{x}\sigma_{i+r}^{y}\rangle$,
shows oscillating decrease. Surprisingly, the coherence measures,
including ${\cal I}(\rho_{i,i+r},\sigma_i^x)$ and
${\cal I}(\rho_{i,i+r},\sigma_i^z)$, exhibit a smooth decay. The first
derivative of the local coherence ${\cal I}(\rho_{ij},\sigma^x_i)$
correctly spotlights the location of the second-order QPT at 
$h_c=\sqrt{4D^2-\gamma^2+1}$ through a divergence, but
no sign of the nontrivial factorization region can be observed.

\begin{figure}[t!]
\noindent
\includegraphics[width=1.02\columnwidth]{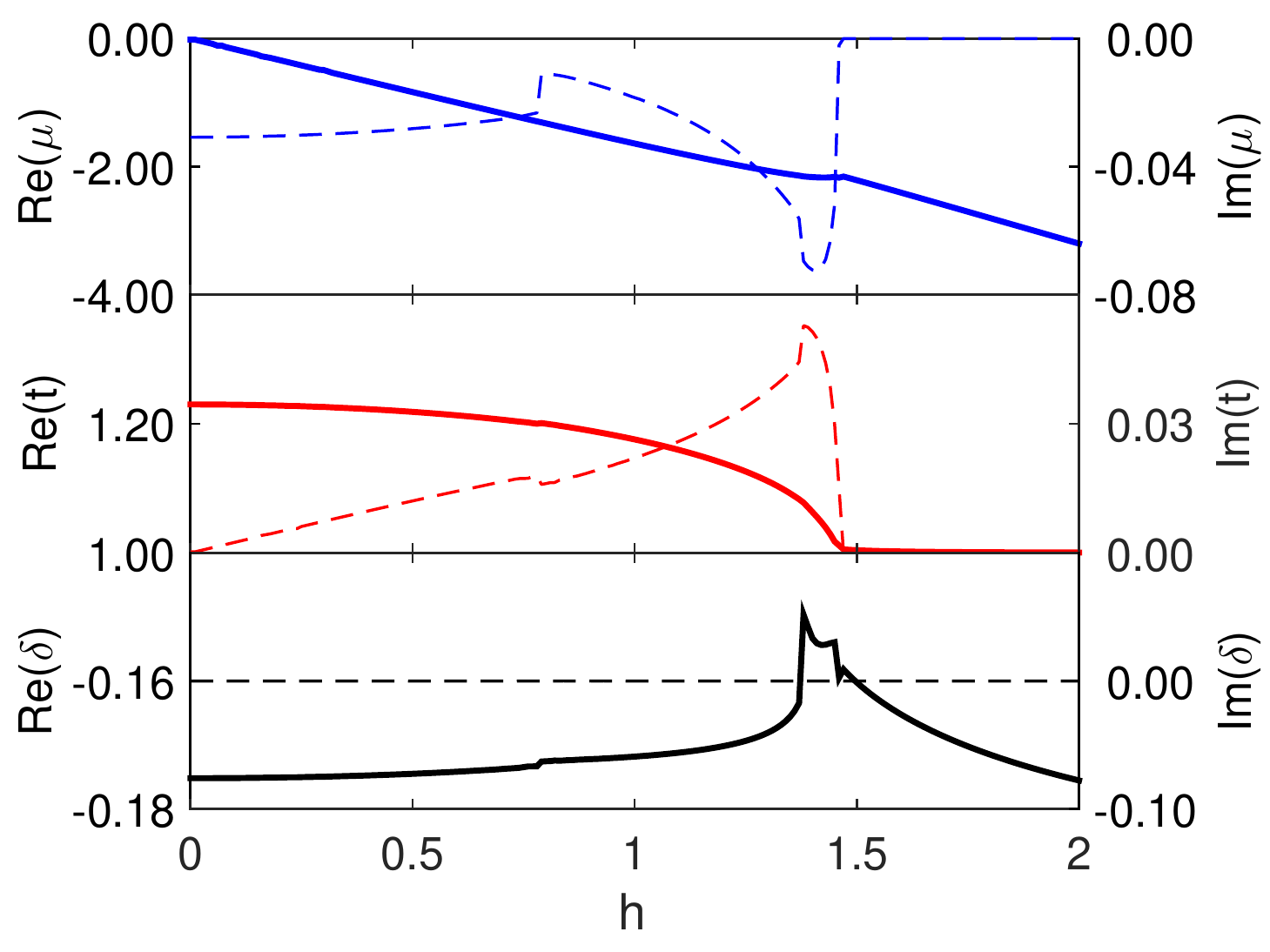}
\caption{Self-consistent mean-field parameters $\mu$, $t$, and $\delta$
as a function of $h$ for $\Delta=0.2$, $D=0.2$, $\gamma=0.2$, and $J=1$.
The solid lines correspond to the real parts, while the dotted lines
represent the imaginary parts.
 }\label{fig:Heisenberg_para}
\end{figure}

\begin{figure}[b!]
\noindent
\includegraphics[width=1.02\columnwidth]{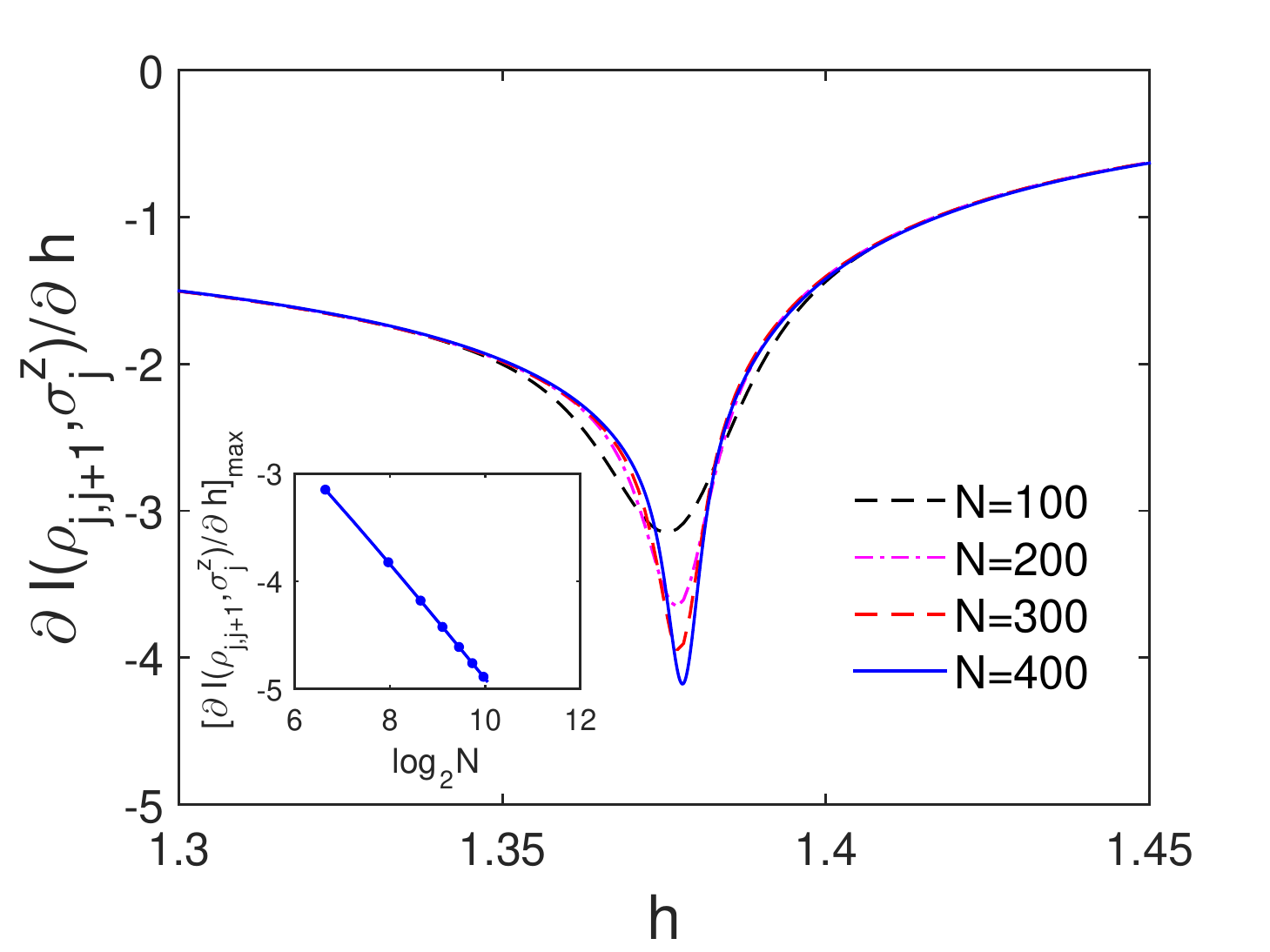}
\caption{The first-derivative of LQC$_z$ with respect to $h$ for
different system sizes with $\gamma$=0.6, $D$=0.2, $\Delta$=0.2, and $J=1$.
Inset shows the relation between maxima of the peaks and the
logarithm of system size $N$.}
\label{fig:Heisenberg_scaling}
\end{figure}

To proceed, we examine the relation between the quantum-information
quantifier with QPTs in the presence of anisotropy term $\Delta$.
For axial regime $\Delta\gg 1$ ($\Delta\ll -1$), the system effectively
stays in the N\'{e}el (ferromagnetic) phase~\cite{Thakur18}.
In the following, we concentrate on the planar regime
($\vert\Delta\vert<1$).
Figure~\ref{fig:Heisenberg_coherence} shows the corresponding behavior
of the entanglement and the quantum coherence with $\gamma=0.6$,
$\Delta=0.2$ and $D=0.2$. One observes that comparing with
Fig.~\ref{fig:entanglement}, the presence of the term $\propto\Delta$
merely increases the values of field at the critical point $h_c$ and
the factorized point $h_f$.
Meanwhile, the hopping parameters $\mu$ and $t$ are found to be
complex, but the pairing parameter $\delta$ remains a real number,
as is disclosed in Fig.~\ref{fig:Heisenberg_para}.

\begin{figure}[t!]
\noindent
\includegraphics[width=1.05\columnwidth]{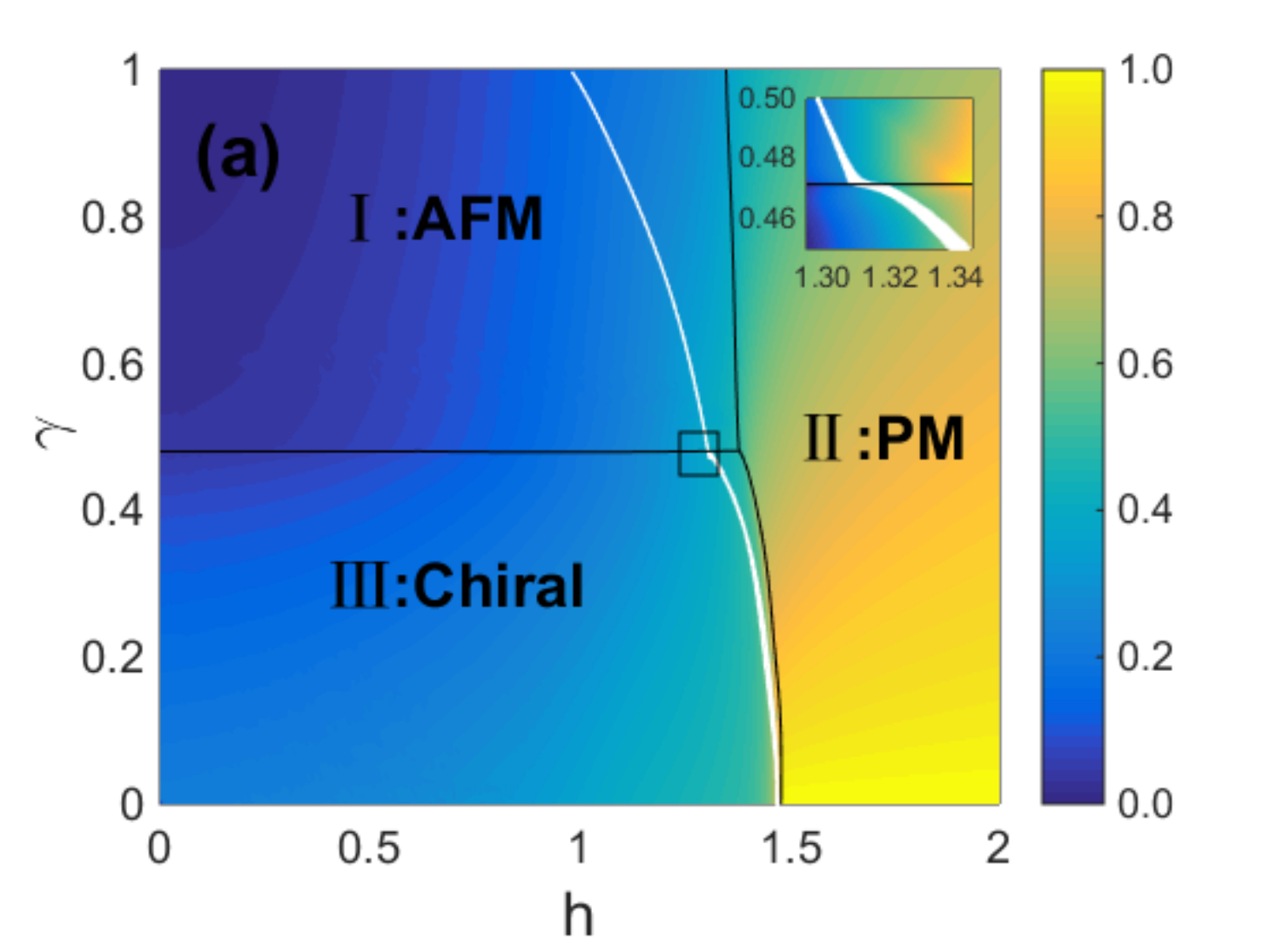}
\includegraphics[width=1.05\columnwidth]{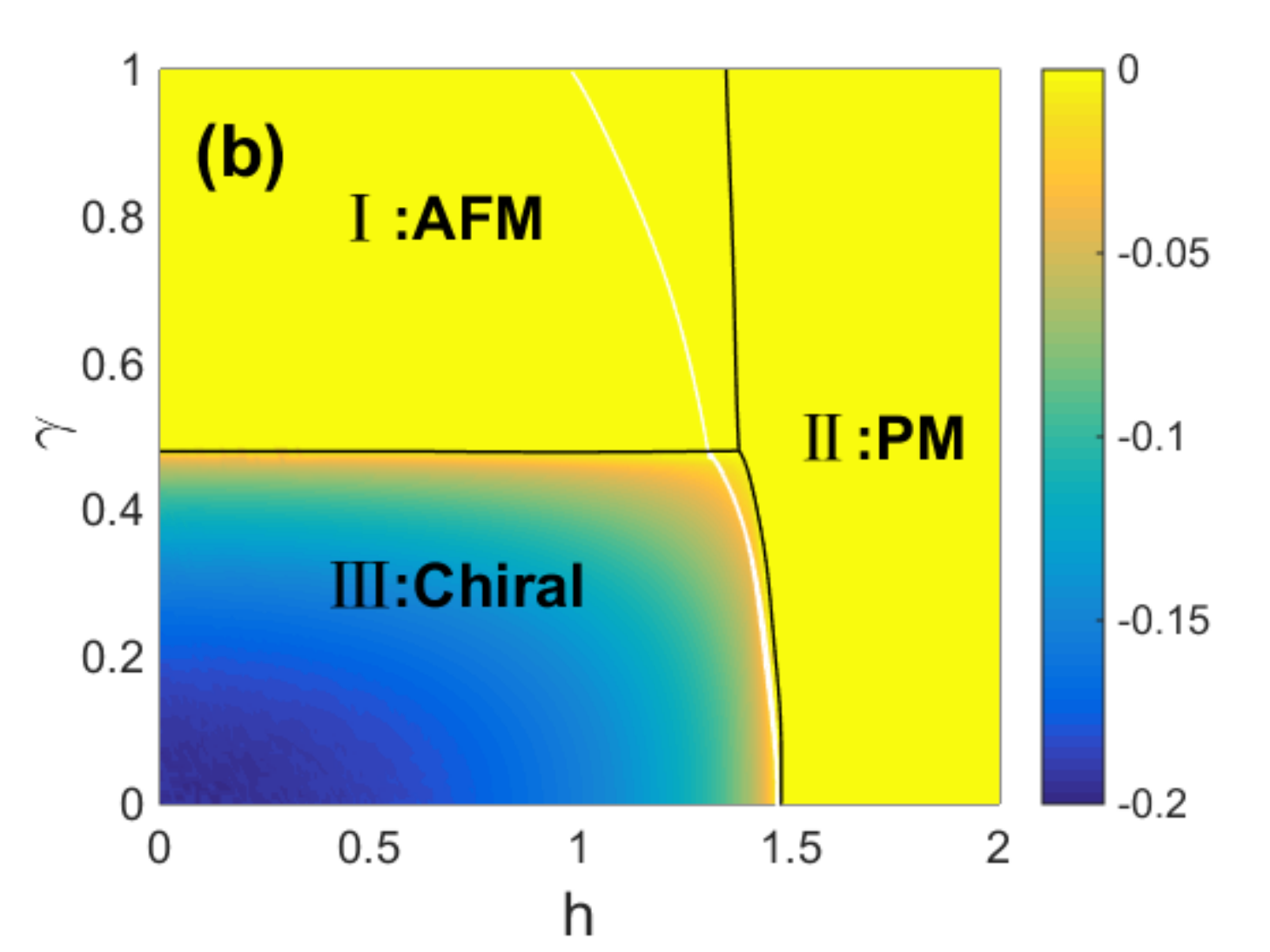}
\caption{
Different parameter regimes of the 1D Heisenberg model in the
$(h,\gamma)$ plane obtained:
(a) phase diagram; the color map represents the strength of the LQC$_x$
${\cal I}(\rho_{i,i+1},\sigma_i^x)$;
(b) two-qubit correlation function
$\langle\sigma_i^x\sigma_{i+1}^y\rangle$; the color map represents the
strength of the correlation function. The white zone in the AFM phase
and the chiral phase delegates the factorization line
and the factorization region, respectively.
Parameters: $D=0.2$ , $\Delta=0.2$, and $J=1$.
 }\label{fig:HeiPhD}
\end{figure}

The first-order derivative of the LQC$_z$ still follows a logarithmic
divergence across the critical point, as is shown in the inset of
Fig.~\ref{fig:Heisenberg_scaling}. The calculations are summarized by
the ground-state phase diagram shown in Fig.~\ref{fig:HeiPhD}(a).
One finds that $\langle\sigma_{i}^{x}\sigma_{i+1}^{y}\rangle=0$ in
the gapped phase [see Fig.~\ref{fig:HeiPhD}(b)], and these three
self-consistent parameters are found to be real numbers as a function
of~$h$. On the other hand, when the system is in the gapless phase,
$\langle\sigma_i^x\sigma_{i+1}^y\rangle$ becomes finite. Such a feature
implies that $\langle\sigma_i^x\sigma_{i+1}^y\rangle$ is an order
parameter to identify the chiral phase for general Heisenberg XYZ
model.

\section{Conclusion and summary}
\label{conclusion}%

In this article, we studied the one-dimensional XYZ model with
Dzyaloshinskii-Moriya interaction, which induces a gapless chiral
phase. We point out a few differences in deriving the exact correlation
functions in this chiral phase and the associated density matrix in
systems with broken reflection symmetry, which then give rise to the
misleading message about the quantum criticality. We firstly scrutinize
the limiting situation, where the XY chain is rigorously solvable by
applying the Jordan-Wigner transformation. Knowledge of exact solutions
endowed with precisely determined properties of separability or
criticality can be of great relevance in the study of general cases,
that are not exactly solvable. For models not admitting exact general
solutions, we carry on an analytical approach that combines a
Jordan-Wigner transformation with a mean-field approximation.
We find the Wigner-Yanase skew information as a quantum coherence
witness which may well identify the quantum phase transitions.

Besides, the logarithmic scaling behavior for the information measures
are found around quantum criticality. Quantum coherence arising from the
quantum superposition acts as one of perspective towards a kaleidoscope
of quantum correlations, and it is the key resource for applications of
quantum technology besides entanglement and other types of quantum
correlations. We have seen that the ground states of complex quantum
systems are typically entangled. Nevertheless, for some specific values
of the parameters, a ground state may be completely separable.

We also discussed the occurrence of the separable ground state in the
antiferromagnetic phase, which is marked by the vanishing of the
concurrence. Such factorization points can be also sensed by the
discontinuous jump of the first derivative of the Wigner-Yanase skew
information measure. In the gapless chiral phase, the factorization
line becomes the factorization volume, which is implied by the
extinguishment of ground-state pairwise concurrence. A merit of the
concurrence and local quantum coherence is that the property emerges
for a finite chain, in contrast to the signal of global entanglement
can be observed in the thermodynamics limit after taking the phase-flip
symmetry breaking into account~\cite{Rad18}. As most multipartite
measures are exhaustively expensive to obtain, the bipartite measures
are comparably easy to calculate, and especially can be determined
without a full tomography of the state. Nevertheless, the vanishing
concurrence is a necessary condition for the occurrence of a completely
separable state, and hence a confirmative conclusion desires further
investigations.

\acknowledgments
W.-L. You acknowledges NSFC under Grant No. 11474211.
N. Wu acknowledges NSFC under Grant No. 11705007. 
A. M. Ole\'s kindly acknowledges support by Narodowe Centrum Nauki
(NCN, Poland) under Project No. 2016/23/B/ST3/00839 and is
\mbox{grateful} for the Alexander von Humboldt Foundation
Fellowship (Humboldt-Forschungspreis).

\appendix

\section{Exact solution of XY chain and correlations}
\label{appendix-Exactsolution}

For $\Delta=0$, the diagonalization procedure of the Heisenberg model
(\ref{eq:ham}) includes the well-established techniques of Jordan-Wigner
and Bogoliubov transformations. We use Jordan-Wigner transformation,
i.e.,
\begin{equation}
\sigma^{+}_{j}\equiv \frac{1}{2}(\sigma^{x}+ i \sigma^{y}) =
e^{i\pi\sum_{n<j}c_{n}^{+}c_{n}^{}}c_{j}^{},
\end{equation}
to covert the spin operators to fermion operators. As a result, the
Hamiltonian (\ref{eq:ham}) can be written as the quadratic form of the
creation operator and the annihilation operator of spinless fermions 
($J$=1 is assumed in the following),
\begin{eqnarray}
    \cal{H}&=& \sum_{j=1}^{N}\left[
     (1+2{i}{D}){c}_{j+1}^{+}{c}_{j}
    + (1-2{i}{D}){c}_{j}^{+}{c}_{j+1}\right. \nonumber \\
&+&\left.{ \gamma}({c}_{j+1}{c}_{j}+{c}_{j}^{+}{c}_{j+1}^{+})
    -{h}(1-2{c}_{j}^{+}{c}_{j})\right].
    \label{Hamiltonian2}
    \end{eqnarray}
In the next step we adopt Fourier transformation to express Eq.
(\ref{Hamiltonian2}) in the momentum space. Then by successive
application of the Bogoliubov transformation, this Hamiltonian
can be reduced to a diagonal form:
\begin{eqnarray}
\cal{H}&=& \sum_{k=-\pi}^{\pi}{\epsilon}_{k}
\left({f}_{k}^{\dagger}{f}_{k}^{}-\frac{1}{2}\right),
\end{eqnarray}
where
 \begin{eqnarray}
{\epsilon}_{k}=-4{D\sin{k}}+2\sqrt{(\cos{k}+{h})^2+(\gamma\sin{k})^2}. 
\quad\label{epsilonk}
\end{eqnarray}

The ground state $\vert\Psi_0\rangle$ follows the total filling of
the Fermi-Dirac statistics, and the lowest energy is obtained when all
the states with negative energies ($\epsilon_k<0$) are filled by
fermions and the ones with positive energies ($\epsilon_k\ge 0$) are
empty. With Eq. (\ref{epsilonk}), the gap $\Delta\equiv\min_k\epsilon_k$
closes at the critical mode $k_{c}$ and the critical field $h_c$, and
then we have
\begin{eqnarray}
h&=&-\cos k+i\sqrt{\gamma^2-4D^2}\,\sin k, \quad\textrm{for}\quad
\gamma> 2D, \nonumber \\
h&=&-\cos k+\,\,\sqrt{4D^2-\gamma^2}\,\sin k, \quad\textrm{for}\quad
\gamma\le 2D. \nonumber
\end{eqnarray}
The reality of $h_c$ requires that $h_c=1$ and $k_c=\pi$ are the
solutions for $\gamma>2D$, while for $\gamma \le 2D$ there are solutions
with arbitrary $k$, suggesting the system is always gapless for
$h<\sqrt{4D^2-\gamma^2+1}$. The analysis suggests the critical lines are
$h=1$, $\gamma=2D$, and \mbox{$h=\sqrt{4D^2-\gamma^2+1}$,} respectively.

The phase diagram at finite DM interaction and finite magnetic field
consists of three phases: AFM phase, PM phase, and the gapless chiral
phase. The transition from AFM phase to PM phase for $\gamma>2D$ is
similar to the conventional order-disorder transition in the transverse
Ising model for $\gamma=1$ and $D=0$. They are in the same universality
class. The AFM phase disappears only when $\gamma=0$. With $D$ getting
smaller, the chiral phase shrinks. When the DM interaction is large
($\gamma<2D$), part of the spectrum becomes negative and the energy gap
disappears, with two Fermi points $k_{\rm L}$ and $k_{\rm R}$ given by
\begin{eqnarray}
k_{\rm L,R}=\cos^{-1}\left[
\frac{-h\pm\sqrt{(4 D^2-\gamma ^2)(4 D^2-\gamma ^2+1-h^2)}}
{4 D^2-\gamma ^2+1}\right].\nonumber \\
\end{eqnarray}
For $k_{\rm L }\le k\le k_{\rm R}$, the excitation spectrum $\epsilon_k$
becomes negative, and these modes in the ground state
$\vert\psi_0 \rangle$ are occupied by electrons, namely
$f_k^\dagger\vert\psi_0\rangle=0$ \cite{tong2013}. The system enters
the gapless chiral phase. As the magnetic field $h$ increases, the
system changes from the chiral phase to the PM phase.

In order to identify different phases, we choose correlation function
between two lattice sites as the order parameter, which can be used to
describe the nature of the ground state.
The correlation function can be defined as:
$G_{i,j}^{\alpha\beta}\equiv
\langle { \sigma}_{i}^{\alpha}{ \sigma}_{j}^{\beta}\rangle -
\langle { \sigma}_{i}^{\alpha}\rangle\langle{ \sigma}_{j}^{\beta}\rangle$,
here $\alpha,\beta=x,y,z$.
Since the system is translation-invariant, the value of the correlation
function is only related to the relative distance between the position
of the two sites (such as $i$ and $j$), so $G_{i,j}^{\alpha\beta}$ can
be abbreviated as $G_{r}^{\alpha\beta}$, here $r=i-j$.
For general ${\langle { \sigma}_{i}^{x}{ \sigma}_{j}^{x}\rangle}$,
${\langle { \sigma}_{i}^{y}{ \sigma}_{j}^{y}\rangle}$,
the expanded form can be expressed as a form of Pfaffian \cite{Bar71}.
In other words, it can be written as the determinant of the $2n\times 2n$
($n\equiv\vert j-i \vert$) dimension anti-symmetric matrix.

It is illuminating to discuss the asymptotic behavior of the
correlation functions in the exact case. Barouch and McCoy studied the
magnetization and the correlation function of XY chain in a transverse
field \cite{Bar70,Bar71}. Its considered the non-zero
temperature correlations of the horizontal-field XX model \cite{Its93}.
The research shows that the asymptotic behavior of the correlation
function($r\to\infty$) can be written in Ornstein-Zernike form specially
in the context of 1D systems significantly away from the critical points
\cite{Landau-Book,Sun16}：
\begin{equation}
 G_{r}^{xx} \sim  A {r^{1-\eta_x}} \exp(-r/\xi),
\end{equation}
where $A$ is a form factor, $\langle\sigma^x_i\rangle$ is the
magnetization in the $x$ direction and $\xi$ is the correlation length.
$\eta_x$ is the Tomonaga-Luttinger exponents for the $x$ spin component
\cite{Bunder99}, which is in the algebraic behavior is equal to $1/2$
for such an asymptotic behavior of the correlation function
\cite{Rams2015}. In all cases $(-1)^r G_{r}^{xx}$ vanishes exponentially
rapidly as $r\to\infty$ for all $h$ and $\gamma$.

However, the rate of this exponential vanishing depends on $h$, and this
dependence is qualitatively different in different regions. When $|h|>1$,
the system is in the paramagnetic state and the magnetization in the $x$
direction disappears, namely, $\langle\sigma^x_i\rangle$=0, and at this 
time ${\lim_{r\to\infty}} G_{r}^{xx}\sim (-1)^r r^{-1/2}\exp(-r/\xi)$,
${\lim_{r \to\infty}} G_{r}^{yy} \sim (-1)^r r^{-3/2} \exp(-r/\xi)$.
When $|h|\le{1}$,
${\lim_{r\to\infty}} G_r^{xx}=
(-1)^r 2[{\gamma}^2(1-h^2)]^{1/4}(1+\gamma)^{-1}$.
This means that when $\gamma \neq 0$, there is a long range order.
When $\gamma = 0 $, the long range order does not exist.
In the Ising limit, ${\lim_{r \to 0}} G_{r}^{xx}=(-1)^r (1-h^2)^{1/4}$.
This implies that the critical exponent $\beta$ is 1/8 for the Ising
transition (approaching the transition as a ferromagnet) and 1/4 for
the anisotropic transition.
$G_r^{xx}$ decreases to zero rapidly with the increasing of $r$ in the
paramagnetic phase, while $G_r^{xx}$ remains a constant with the change
of $r$ in the AFM phase.

At the critical point of Ising transition ($h=1$),
$G_r^{xx}\sim r^{-1/4}$, the critical exponent $\eta_x=5/4$.
$G_r^{yy}\sim r^{-9/4}$ with $\eta_y=13/4$. At the anisotropic phase
transition line when $h=0$ and $\gamma=0$, $G_r^{xx}\sim r^{-1/2}$
\cite{Bunder99}.
The DM interactions cause the correlation function $G_r^{xy}$ to
decrease oscillatory with the increase of the distance $r$. When
$\gamma=0$, the correlation function $G_r^{xy}$ oscillates more
violently than $\gamma=1$ with the increase of the distance $r$.

For specific values of the anisotropy parameter and the relative
strengths of the uniform transverse magnetic fields, the ground state
of this model is known to be doubly degenerate and factorizable along
two hyperbolic lines, known as the factorization lines. For the
factorization points $h^2+\gamma^2=1$ with $D=0$, we can obtain an
explicit form for all $r$:
$\langle\sigma_j^x\sigma_{j+r}^x\rangle=(-1)^r 2 \gamma/(1+\gamma)$,
$\langle\sigma_j^y\sigma_{j+r}^y\rangle=0$,
$\langle\sigma_j^z\sigma_{j+r}^z\rangle=\langle\sigma_j^z\rangle^2$.

\newpage
\section{Fermionic mean-field approximation}
\label{appendix-MF}
According to the mean-field decomposition in Eq. (\ref{eqMF}),
the order parameters are defined as
\begin{eqnarray}
\beta_1&=&\langle{c^+_{j}c_{j}}\rangle ,   \\
\beta_2&=&\langle{c^+_{j}c_{j+1}}\rangle ,   \\
\beta_3&=&\langle{c_{j}c_{j+1}}\rangle.
\end{eqnarray}

The energy spectrum (\ref{epsilonk}) can be rewritten as
\begin{eqnarray}
\varepsilon(k) &=&  -4D\sin(k)+2\sqrt{\tau(k)^2+\varphi(k)^2},
\end{eqnarray}
where $\tau(k)=J[(1-4\Delta\beta_2)\cos(k)+2\Delta(2\beta_1-1)]+h$
and $\varphi(k)=J(\gamma-4\Delta\beta_3)\sin(k)$.
To this end, one finds the solutions could be retrieved by
self-consistently solving the following equations:
\begin{eqnarray}
\beta_1&=&\sum_{k=-\pi}^{\pi}[
\frac{\varphi(k)^2\theta(-\varepsilon_k)+\varsigma(k)^2
\theta(\varepsilon_{-k})}{\varphi(k)^2+\varsigma(k)^2}, \\
\beta_2&=&\sum_{k=-\pi}^{\pi}[
\frac{\varphi(k)^2\theta(-\varepsilon_k)+\varsigma(k)^2
\theta(\varepsilon_{-k})}{\varphi(k)^2+\varsigma(k)^2}e^{-ik},\\
\beta_3&=&\sum_{k=-\pi}^{\pi}[
\frac{-\theta(-\varepsilon_k)+\theta(\varepsilon_{-k})}
{\varphi(k)^2+\varsigma(k)^2}(i\varphi(k)\varsigma(k)) e^{ik},
\label{self}
\end{eqnarray}
with $\varsigma(k)=\tau(k)+\sqrt{\tau(k)^2+\varphi(k)^2}$.


\begin{thebibliography}{1}

\bibitem{Sachdev1999} S. Sachdev,
                   \textit{Quantum Phase Transition}
                   (Cambridge University Press, Cambridge, England, 1999).

\bibitem{hwang2015prl} M.-J. Hwang, R. Puebla, and M. B. Plenio,
                   Phys. Rev. Lett. \textbf{115}, 180404 (2015).

\bibitem{hwang2016prl} M.-J. Hwang and M. B. Plenio,
                   Phys. Rev. Lett. \textbf{117}, 123602 (2016).

\bibitem{Liu17} M. Liu, S. Chesi, Z.-J. Ying, X. Chen, H.-G. Luo, and H.-Q.Lin,
                   Phys. Rev. Lett. \textbf{119}, 220601 (2017).

\bibitem{Lar17} J. Larson and E. K. Irish,
                   J. Phys. A: Math. Theor. \textbf{50}, 174002 (2017).

\bibitem{wang2018njp} Y. M. Wang, W.-L. You, M. X. Liu, Y.-L. Dong, 
                   \mbox{H.-G.} Luo, G. Romero, and J. Q. You,
                   New J. Phys. \textbf{20}, 053061 (2018).

\bibitem{Aberg2014} J. {\AA}berg,
                   Phys. Rev. Lett. \textbf{113}, 150402 (2014).

\bibitem{Narasimhachar2015} V. Narasimhachar and G. Gour,
                   Nat. Commun. \textbf{6}, 7689 (2015).

\bibitem{Cwi15} P. \'{C}wikli\'nski, M. Studzi\'nski, M. Horodecki, 
                   and \mbox{J. Oppenheim,}
                   Phys. Rev. Lett. \textbf{115}, 210403 (2015).

\bibitem{Lo15N} M. Lostaglio, D. Jennings, and T. Rudolph,
                   Nature Commun. \textbf{6}, 6383 (2015).

\bibitem{Lo15X} M. Lostaglio, K. Korzekwa, D. Jennings, and T.~Rudolph,
                   Phys. Rev. X \textbf{5}, 021001 (2015).

\bibitem{Schrodinger} E. Schr\"{o}dinger,
                   Math. Proc. Cambridge Philos. Soc. \textbf{31}, 555 (1935).

\bibitem{You15} W.-L. You, A. M. Ole\'s, and P. Horsch,
                   New J. Phys. \textbf{17}, 083009 (2015).

\bibitem{Baumgratzq14} T. Baumgratz, M. Cramer, and M. B. Plenio,
                   Phys. Rev. Lett. \textbf{113}, 140401 (2014).

\bibitem{Str15} A. Streltsov, U. Singh, H. S. Dhar, M. N. Bera, and G.~Adesso,
                   Phys. Rev. Lett. \textbf{115}, 020403 (2015).

\bibitem{Braun18} D. Braun, G. Adesso, F. Benatti, R. Floreanini,
                   \mbox{U. Marzolino,} M. W. Mitchell, and S. Pirandola,
                   Rev. Mod. Phys. \textbf{90}, 035006 (2018).

\bibitem{Chen16} J.-J. Chen, J. Cui, Y.-R. Zhang, and H. Fan,
                   Phys. Rev. A \textbf{94}, 022112 (2016).

\bibitem{You17} W.-L. You, C.-J. Zhang, W. Ni, M. Gong, and A.~M.~Ole\'s,
                   Phys. Rev. B \textbf{95}, 224404 (2017).

\bibitem{Zanardi07} P. Zanardi, H. T. Quan, X. Wang, and C. P. Sun,
                   Phys. Rev. A \textbf{75}, 032109 (2007).

\bibitem{You07} W.-L. You, Y.-W. Li, and S.-J. Gu,
                   Phys. Rev. E \textbf{76} 022101 (2007).

\bibitem{Kur82} J. Kurmann, H. Thomas, and G. M\"{u}ller,
                   Physica A \textbf{112}, 235 (1982).

\bibitem{Horodecki09} R. Horodecki, P. Horodecki, M. Horodecki, and K.~Horodecki,
                   Rev. Mod. Phys. \textbf{81}, 865 (2009).

\bibitem{Vandersypen05}L. M. K. Vandersypen and I. L. Chuang,
                   Rev. Mod. Phys. \textbf{76}, 1037 (2005).

\bibitem{Georgescu14} I. M. Georgescu, S. Ashhab, and F. Nori,
                   Rev. Mod. Phys. \textbf{86}, 153 (2014).

\bibitem{You14} W.-L. You, G.-H. Liu, P. Horsch, and A. M. Ole\'s,
                   Phys. Rev. B \textbf{90}, 094413 (2014).

\bibitem{Thakur18} P. Thakur and P. Durganandini,
                   Phys. Rev. B \textbf{97}, 064413 (2018).

\bibitem{Brockmann2013} M. Brockmann, A. Kl\"{u}mper, and V. Ohanyan,
                   Phys. Rev. B \textbf{87}, 054407 (2013).

\bibitem{Rad17} C. Radhakrishnan, I. Ermakov, and T. Byrnes,
                   Phys. Rev. A \textbf{96}, 012341 (2017).

\bibitem{Rad16} C. Radhakrishnan, M. Parthasarathy, S. Jambulingam, and T. Byrnes,
                   Phys. Rev. Lett. \textbf{116}, 150504 (2016).

\bibitem{Li2018} S.-P. Li and Z.-H. Sun,
                   Phys. Rev. A \textbf{98}, 022317 (2018).

\bibitem{Porras04} D. Porras and J. I. Cirac,
                   Phys. Rev. Lett. \textbf{92}, 207901 (2004).

\bibitem{Berloff17} N. G. Berloff, M. Silva, K. Kalinin, A. Askitopoulos,
                   J. D. T\"{o}pfer, P. Cilibrizzi, W. Langbein,
                   and P.~G.~Lagoudakis, Nature Mater. \textbf{16}, 1120 (2017).

\bibitem{Dzy58} I. Dzyaloshinskii,
                   J. Phys. Chem. Solids \textbf{4}, 241 (1958).

\bibitem{Mor60} T. Moriya,
                   Phys. Rev. \textbf{120}, 91 (1960).

\bibitem{Seki12} S. Seki, X. Z. Yu, S. Ishiwata, and Y. Tokura,
                   Science \textbf{336}, 198 (2012).

\bibitem{Adams12} T. Adams, A. Chacon, M. Wagner, A. Bauer, G. Brandl,
                   B. Pedersen, H. Berger, P. Lemmens, and C. Pfleiderer,
                   Phys. Rev. Lett. \textbf{108}, 237204 (2012).

\bibitem{Yang12} J. H. Yang, Z. L. Li, X. Z. Lu, M.-H. Whangbo, \mbox{S.-H. Wei,}
                   X. G. Gong, and H. J. Xiang,
                   Phys. Rev. Lett. \textbf{109}, 107203 (2012).

\bibitem{Matsuda} M. Matsuda, R. S. Fishman, T. Hong, C. H. Lee, T.~Ushiyama,
                   Y. Yanagisawa, Y. Tomioka, and \mbox{T. Ito,}
                   Phys. Rev. Lett. \textbf{109}, 067205 (2012).

\bibitem{Wang12} P. Wang, Z.-Q. Yu, Z. Fu, J. Miao, L. Huang, S. Chai,
                   H. Zhai, and J. Zhang,
                   Phys. Rev. Lett. \textbf{109}, 095301 (2012).

\bibitem{Cheuk12} L. W. Cheuk, A. T. Sommer, Z. Hadzibabic, T. Yefsah,
                   W. S. Bakr, and M. W. Zwierlein,
                   Phys. Rev. Lett. \textbf{109}, 095302 (2012).

\bibitem{Garcia12} K. Jim\'enez-Garcia, L. J. LeBlanc, R. A. Williams,
                   M. C. Beeler, A. R. Perry, and I. B. Spielman,
                   Phys. Rev. Lett. \textbf{108}, 225303 (2012).

\bibitem{Cole12} W. S. Cole, S. Zhang, A. Paramekanti, and N. Trivedi,
                   Phys. Rev. Lett. \textbf{109}, 085302 (2012).

\bibitem{Tokura10} Y. Tokura and S. Seki,
                   Adv. Mater. \textbf{22}, 1554 (2010).

\bibitem{Sergienko06} I. A. Sergienko and E. Dagotto,
                   Phys. Rev. B  \textbf{73}, 094434 (2006).

\bibitem{Shekhtman92} L. Shekhtman, O. Entin-Wohlman, and A. Aharony,
                   Phys. Rev. Lett. \textbf{69}, 836 (1992).

\bibitem{Bethe1931} H. A. Bethe,
                   Z. Phys. \textbf{71}, 205 (1931).

\bibitem{Yang1966} C. N. Yang and C. P. Yang,
                   Phys. Rev. \textbf{150}, 321 (1966).

\bibitem{Vionnet2017} G. Vionnet, B. Kumar, and F. Mila,
                   Phys. Rev. B \textbf{95}, 174404 (2017).

\bibitem{juan1} J. Carrasquilla and R. Melko,
                   Nat. Phys. \textbf{13}, 431 (2017).

\bibitem{un1}   L. Wang,
                   Phys. Rev. B \textbf{94}, 195105 (2016).

\bibitem{un2}   S. J. Wetzel,
                   Phys. Rev. E \textbf{96}, 022140 (2017).

\bibitem{Sca17} W. Hu, R. R. P. Singh, and R. T. Scalettar,
                   Phys. Rev. E {\bf 95}, 062122 (2017).


\bibitem{zhai17} C. Wang and H. Zhai,
                   Phys. Rev. B {\bf 96}, 144432 (2017).

\bibitem{beach18} M. J. S. Beach, A. Golubeva, and R. G. Melko,
                   Phys. Rev. B {\bf 97}, 045207 (2018).

\bibitem{zhai18} C. Wang and H. Zhai,
                   Frontiers of Physics {\bf 13}, 130507 (2018).

\bibitem{Zha18} W. Zhang, J. Liu, and T.-C. Wei,
                   Phys. Rev. E {\bf 99}, 032142 (2019).

\bibitem{Rad18} R. Radgohar and A. Montakhab,
                   Phys. Rev. B \textbf{97}, 024434 (2018).

\bibitem{Toskovic16} R. Toskovic, R. van den Berg, A. Spinelli, 
                   I. S. Eliens, \mbox{B. van den Toorn}, B. Bryant, 
                   J. S. Caux, and A. F. Otte,
                   Nature Phys. \textbf{12}, 656 (2016).

\bibitem{Zheng2000} S.-B. Zheng and G.-C. Guo,
                   Phys. Rev. Lett. \textbf{85}, 2392 (2000).

\bibitem{Christandl04} M. Christandl, N. Datta, A. Ekert, and A. J. Landahl,
                   Phys. Rev. Lett. \textbf{92}, 187902 (2004).

\bibitem{Bose2003} S. Bose,
                   Phys. Rev. Lett. \textbf{91}, 207901 (2003).

\bibitem{Giampaolo08} S. M. Giampaolo, G. Adesso, and F. Illuminati,
                   Phys. Rev. Lett. \textbf{100}, 197201 (2008).

\bibitem{Ye2018} B.-L. Ye, B. Li, Z.-X. Wang, X. Li-Jost, and S.-M. Fei,
                   Sci. China-Phys. Mech. Astron. \textbf{61}, 110312 (2018).

\bibitem{Liu11} B. Q. Liu, B. Shao, J. G. Li, J. Zou, and L. A. Wu,
                   Phys. Rev. A \textbf{83}, 052112 (2011).

\bibitem{Li2009} Y.-C. Li and S.-S. Li,
                   Phys. Rev. A \textbf{79}, 032338 (2009).

\bibitem{Liu2016} D. Lu, T. Xin, N. K. Yu, Z. F. Ji, J. X. Chen, G.~L.~Long,
                   J. Baugh, X. H. Peng, B. Zeng, and R. Laflamme,
                   Phys. Rev. Lett. \textbf{116}, 230501 (2016).

\bibitem{You16} W.-L. You, Y.-C. Qiu, and A. M. Ole\'s,
                   Phys. Rev. B \textbf{93}, 214417 (2016).

\bibitem{Lei2015} S. Lei and P. Tong,
                   Physica B \textbf{463}, 1 (2015).

\bibitem{Winger1963} E. P. Wigner and M. M. Yanase,
                   Proc. Natl. Acad. Sci. (U.S.A.) \textbf{49}, 910 (1963).

\bibitem{Gir14} D. Girolami,
                   Phys. Rev. Lett. \textbf{113}, 170401 (2014).

\bibitem{Luo2012} S. L. Luo, S. S. Fu, C. H. Oh,
                   Phys. Rev. A \textbf{85}, 032117 (2012).


\bibitem{Karpat2014} G. Karpat, B. Cakmak, and F. F. Fanchini,
                   Phys. Rev. B \textbf{90}, 104431 (2014).

\bibitem{Wootters} W. K. Wootters,
                   Phys. Rev. Lett. \textbf{80}, 2245 (1998).

\bibitem{Werlang09} T. Werlang, S. Souza, F. F. Fanchini, 
                   and C. J. Villas Boas,
                   Phys. Rev. A \textbf{80}, 024103 (2009).

\bibitem{Yi2018} T.-C. Yi, Y.-R. Ding, J. Ren, Y.-M. Wang, W.-L. You,
                   Acta Phys. Sin. \textbf{67}, 140303 (2018).

\bibitem{Fan18} M.-L. Hu, X.-Y. Hu, J.-C. Wang, Y. Peng, Y.-R. Zhang, 
                   and H. Fan,
                   Phys. Rep. \textbf{07}, 004 (2018).

\bibitem{Woo82} W. K. Wooters and W. H. Zurek,
                   Nature (London) \textbf{299}, 802 (1982).

\bibitem{Ost02} A. Osterloh, L. Amico, G. Falci, and R. Fazio,
                   Nature (London) \textbf{416}, 608 (2002).

\bibitem{Osborne2002} T. J. Osborne and M. A. Nielsen,
                   Phys. Rev. A \textbf{66}, 032110 (2002).

\bibitem{Gu2003} S.-J. Gu, H.-Q. Lin, and Y.-Q. Li,
                   Phys. Rev. A \textbf{68}, 042330 (2003).

\bibitem{Vidal2003} G. Vidal, J. I. Latorre, E. Rico, and A. Kitaev,
                   Phys. Rev. Lett. \textbf{90}, 227902 (2003).

\bibitem{Ollivier2001} H. Ollivier and W. H. Zurek,
                   Phys. Rev. Lett. \textbf{88}, 017901 (2001).

\bibitem{Barber83} M. N. Barber,
                   in: \textit{Phase Transitions and Critical Phenomena}
                   (Academic, London, 1983), Vol. \textbf{8}, 146-259.

\bibitem{Zhu06} S. L. Zhu,
                   Phys. Rev. Lett. \textbf{96}, 077206 (2006).

\bibitem{Tong16} S. G. Lei and P .Q. Tong,
                   Quantum Inf. Process \textbf{15}, 1811 (2016).

\bibitem{Campbell13} S. Campbell, J. Richens, N. L. Gullo, and T. Busch,
                   Phys. Rev. A \textbf{88}, 062305 (2013).

\bibitem{Wei2005} T.-C. Wei, D. Das, S. Mukhopadyay, S. Vishveshwara, 
                   and P. M. Goldbart,
                   Phys. Rev. A \textbf{71}, 060305(R) (2005).

\bibitem{Muller1985} G. M\"{u}ller and R. E. Shrock,
                   Phys. Rev. B \textbf{32}, 5845 (1985).

\bibitem{Roscilde05} T. Roscilde, P. Verrucchi, A. Fubini, S. Haas, 
                   and V.~Tognetti,
                   Phys. Rev. Lett. \textbf{94}, 147208 (2005).

\bibitem{Cerezo16} M. Cerezo, R. Rossignoli, and N. Canosa,
                   Phys. Rev. A \textbf{94}, 042335 (2016).

\bibitem{Cerezo17} M. Cerezo, R. Rossignoli, N. Canosa, and E. R\'{\i}os,
                   Phys. Rev. Lett. \textbf{119}, 220605 (2017).

\bibitem{Bar70} E. Barouch and B. M. McCoy,
                   Phys. Rev. A \textbf{2}, 1075 (1970).

\bibitem{tong2013} M. Zhong, H. Xu, X. X. Liu, and P. Q. Tong,
                   Chin. Phys. B \textbf{22}, 090313 (2013).

\bibitem{Bar71} E. Barouch and B. M. McCoy,
                   Phys. Rev. A \textbf{3}, 786 (1971).

\bibitem{Its93} A. R. Its, A. G. Izergin, V. E. Korepin, and N.~A.~Slavnov,
                   Phys. Rev. Lett. \textbf{70}, 1704 (1993).

\bibitem{Sun16} T. Vekua and G. Sun,
                   Phys. Rev. B \textbf{94}, 014417 (2016).

\bibitem{Landau-Book} D. P. Landau and K. Binder,
                   \textit{A Guide to Monte Carlo Simulations in Statistical Physics}
                   (Cambridge University Press, 2005).


\bibitem{Bunder99} J. E. Bunder and R. H. McKenzie,
                   Phys. Rev. B \textbf{60}, 344 (1999).

\bibitem{Rams2015} M. M. Rams, V. Zauner, M. Bal, J. Haegeman, 
                   and F.~Verstraete,
                   Phys. Rev. B \textbf{92}, 235150 (2015).

\end{thebibliography}
\end{document}